\documentclass[12pt,a4paper]{article}
\pdfoutput=1
\usepackage{epsfig}
\usepackage{comment}
\usepackage{latexsym}
\usepackage{array}
\usepackage{color}
\usepackage{amsmath}
\usepackage{hyperref}
\usepackage[english]{babel}
\usepackage [applemac] {inputenc} 

\newcommand{\mysquare}[0]{\raise-.2ex\hbox{{\Large$\Box$}}}
\def\lsim{\mathrel{\rlap {\raise.5ex\hbox{$ < $}}
{\lower.5ex\hbox{$\sim$}}}}
\def\gsim{\mathrel{\rlap {\raise.5ex\hbox{$ > $}}
{\lower.5ex\hbox{$\sim$}}}} \topmargin -1.5cm \textheight=22.5cm \textwidth=16.5cm
\catcode`\@=11
\newcount\hour
\newcount\minute
\newtoks\amorpm
\hour=\time\divide\hour by60 \minute=\time{\multiply\hour by60 \global\advance\minute by-\hour}
\edef\standardtime{{\ifnum\hour<12 \global\amorpm={am}%
        \else\global\amorpm={pm}\advance\hour by-12 \fi
        \ifnum\hour=0 \hour=12 \fi
        \number\hour:\ifnum\minute<10 0\fi\number\minute\the\amorpm}}
\edef\militarytime{\number\hour:\ifnum\minute<10 0\fi\number\minute}
\def\draftlabel#1{{\@bsphack\if@filesw {\let\thepage\relax
   \xdef\@gtempa{\write\@auxout{\string
      \newlabel{#1}{{\@currentlabel}{\thepage}}}}}\@gtempa
   \if@nobreak \ifvmode\nobreak\fi\fi\fi\@esphack}
        \gdef\@eqnlabel{#1}}
\def\@eqnlabel{}
\def\@vacuum{}
\def\draftmarginnote#1{\marginpar{\raggedright\scriptsize\tt#1}}
\def\draft{\oddsidemargin -.2truein
        \def\@oddfoot{\sl preliminary draft \hfil
        \rm\thepage\hfil\sl\today\quad\militarytime}
        \let\@evenfoot\@oddfoot \overfullrule 3pt
        \let\label=\draftlabel
        \let\marginnote=\draftmarginnote
   \def\@eqnnum{(\theequation)\rlap{\kern\marginparsep\tt\@eqnlabel}
\global\let\@eqnlabel\@vacuum}  }

\newcommand{\be}[0]{\begin{equation}}
\newcommand{\ee}[0]{\end{equation}}
\newcommand{\ba}[0]{\begin{eqnarray}}
\newcommand{\ea}[0]{\end{eqnarray}}

\def\bs{\begin{subequations}}
\def\es{\end{subequations}}

\def\thebibliography#1{%
\vskip 0.5cm \centerline{\bf \Large References}
\list{%
[\arabic{enumi}]}{\settowidth\labelwidth{[#1]} \leftmargin\labelwidth \advance\leftmargin\labelsep
\usecounter{enumi}}
\def\newblock{\hskip .11em plus .33em minus .07em}
\sloppy\clubpenalty4000\widowpenalty4000 \sfcode`\.=1000\relax}

\renewcommand{\theequation}{\arabic{section}.\arabic{equation}}

\renewcommand{\section}{\setcounter{equation}{0}\@startsection
{section}{1}{0mm}{-\baselineskip}{0.5\baselineskip} {\normalfont\Large\bfseries}}

\renewcommand{\subsection}{\@startsection
{subsection}{2}{0mm}{-\baselineskip}{0.5\baselineskip} {\normalfont\large\bfseries}}

\renewcommand{\subsubsection}{\@startsection
{subsubsection}{3}{0mm}{-\baselineskip}{0.5\baselineskip} {\normalfont\normalsize\slshape}}

\usepackage{amssymb,amsfonts}
\usepackage{graphicx}
\usepackage{cite}

\newcommand{\bea}{\begin{eqnarray}}
\newcommand{\eea}{\end{eqnarray}}
\newcommand{\dis}{\displaystyle}

\newcommand{\esp}{\!\!\!\phantom{\Bigg\abs}}

\newcommand{\CP}{\mathbb{P}}
\newcommand{\Z}{\mathbb{Z}}
\newcommand{\Ka}{K{\"a}hler }
\renewcommand{\O}{{\cal O}}

\newcommand{\abs}{|}
\newcommand{\dprime}{{\prime\prime}}
\newcommand{\tr}{{\rm tr\,}}

\newcommand{\ie}{{\em i.e. }}
\newcommand{\where}{\mbox{where}}

\renewcommand{\and}{\mbox{and}}

\newcommand{\F}{{\cal F}}
\newcommand{\R}{{\cal R}}
\newcommand{\N}{{\cal N}}
\newcommand{\A}{{\cal A}}
\newcommand{\B}{{\cal B}}
\newcommand{\C}{{\cal C}}
\newcommand{\K}{{\cal K}}
\newcommand{\T}{{\cal T}}
\newcommand{\V}{{\cal V}}
\renewcommand{\P}{{\cal P}}
\newcommand{\U}{{\cal U}}
\newcommand{\M}{{\cal M}}

\renewcommand{\L}{{\cal L}}

\newcommand{\cC}{{\mfs C}}

\newcommand{\cP}{{\mfs P}}

\renewcommand{\b}{\bar}
\newcommand{\h}{\hat}

\renewcommand{\t}{\tilde}

\def\no{\nonumber}
\newcommand{\half}{{1 \over 2}}

\def\ca{{\cal A}}
\def\cb{{\cal B}}

\def\cf{{\cal F}}

\def\cm{{\cal M}}

\newcommand{\demi}{\frac{1}{2}}

\def\ha{{\hat a}}

\DeclareFontFamily{U}{rsfs}{}         
\DeclareFontShape{U}{rsfs}{m}{n}{<5> rsfs5 <6><7> rsfs7          %
  <8><9><10><10.95><12><14.4><17.28><20.74><24.88> rsfs10}{}     %
\DeclareMathAlphabet{\mathfs}{U}{rsfs}{m}{n}                     %
\newcommand{\mfs}[1]{\mathfs {#1}}                               %

\begin{document}
\begin{titlepage}
\begin{flushright}
CPHT--RR098.1111, November 2011
\end{flushright}

\vspace{5mm}

\begin{centering}

{\bf \Large Moduli stabilization in type II Calabi-Yau\\
\vspace{3mm}
compactifications at finite temperature}

\vspace{6mm}
 {\Large Lihui Liu and Herv\'e Partouche}

\vspace{4mm}

Centre de Physique Th\'eorique, Ecole Polytechnique,$^\dag$\\
F--91128 Palaiseau cedex, France

{\em lihui.liu@cpht.polytechnique.fr}\\
{\em herve.partouche@cpht.polytechnique.fr}
\vspace{8mm}

{\bf\Large Abstract}

\end{centering}
\vspace{4mm}

\noindent We consider the type II superstring  compactified on Calabi-Yau threefolds, at finite temperature. The latter is implemented at the string level by a free action on the Euclidean time circle. We show that all \Ka and complex structure moduli involved in the gauge theories geometrically engineered in the vicinity of singular loci are lifted by the stringy thermal effective potential. The analysis is based on the effective gauged supergravity at low energy, without integrating out the non-perturbative BPS states becoming massless at the singular loci. The universal form of the action in the weak coupling regime and at low enough temperature is determined in two cases. Namely, the conifold locus, as well as a locus where the internal space develops a genus-$g$ curve of $A_{N-1}$ singularities, thus realizing an $SU(N)$ gauge theory coupled to $g$ hypermultiplets in the adjoint. In general, we argue that the favored points of stabilization sit at the intersection of several such loci. As a result, the entire vector multiplet moduli space is expected to be lifted, together with hypermultiplet moduli.
The scalars are dynamically stabilized during the cosmological evolution induced by the back-reaction of the thermal effective potential on the originally static background. When the universe expands and the temperature $T$ drops, the scalars converge to minima, with damped oscillations.  Moreover, they store an energy density that scales as $T^4$, which never dominates over radiation. The reason for this is that the mass they acquire at one-loop is of order the temperature scale, which is time-dependent rather than constant.
As an example, we analyze the type IIA compactification on a hypersurface $\CP^4_{(1,1,2,2,6)}[12]$, with Hodge numbers $h_{11}=2$ and $h_{12}=128$. In this case, both \Ka moduli are stabilized at a point, where the internal space develops a node and an enhanced $SU(2)$ gauge theory coupled to 2 adjoint hypermultiplets. This shows that in the dual thermal heterotic picture on $K3\times T^2$, the torus modulus and the axio-dilaton are stabilized, though in a strong coupling heterotic regime.

\vspace{3pt} \vfill \hrule width 6.7cm \vskip.1mm{\small \small \small
\noindent
$^\dag$\ Unit{\'e} mixte du CNRS et de l'Ecole Polytechnique,
UMR 7644.}

\end{titlepage}
\newpage
\setcounter{footnote}{0}
\renewcommand{\thefootnote}{\arabic{footnote}}
 \setlength{\baselineskip}{.7cm} \setlength{\parskip}{.2cm}

\setcounter{section}{0}


\section{Introduction}
\label{intro}

The presence of moduli fields in supersymmetric compactifications of string theory leads to difficulties. Massless scalars are not only in contradiction with observations of the gravitational force (see for example \cite{Adelberger:2003zx}), they also lead to continuous parameters in the couplings and mass spectrum, implying a loss of predictability of the theory. Moreover, moduli spaces which are nothing but the flat directions of a scalar potential often admit particular loci, where states generically massive become massless.
In the literature, several mechanisms indicate these loci correspond to dynamically preferred values of the scalar vacuum expectations values (VEV's). In the context of M-theory or type II compactifications on Calabi-Yau (CY) spaces, it is shown in \cite{eran} that if the scalars are given initial conditions away from the minima of the potential, their temporal trajectories are attracted toward the loci of additional massless states. In \cite{Kofman:2004yc}, the scalar initial conditions are set along the flat directions, but with non-trivial velocities. The moduli motion induces particle productions, whose back-reaction implies again an attraction toward the same loci. However, if the scalars are initially along their flat directions and static, the above mechanisms are ineffective.  Moreover, even in the cases they manage to dynamically select expectation values, the moduli fields remain massless at the end of the process, and additional massless scalars may even be present at these particular points.

On the contrary, the existence of flat directions in non-supersymetric theories is much more sparse \cite{min,Nair:1986zn}. To avoid the presence of a very large cosmological constant, it is natural to focus on ``no-scale models", which by definition are tree level backgrounds in Minkowski space, where supersymmetry is spontaneously broken \cite{Noscale}. If at the classical level the scale of supersymmetry breaking and other scalars are moduli fields, the associated flat directions are generically lifted at the quantum level, due to the generation of a non-trivial effective potential.
In fact, any supersymmetric string compactification in flat space can lead to a no-scale model by switching on finite temperature. This can be done at the level of the conformal field theory on the worldsheet by compactifying the Euclidean time on a circle and modding out by the $\Z_2$ freely acting orbifold $(-1)^{F} \, \delta$, where $F$ is the fermion number and $\delta$ is an order-two shift along the temporal circle. Physically, this is equivalent to imposing  $(-1)^F$ boundary conditions along an Euclidean circle of perimeter equal to the inverse temperature \cite{SS,KoR}. In this case, the supersymmetry breaking scale is the temperature itself, while the effective potential is nothing but the free energy density $\F$.

The question of moduli stabilization in a universe filled with a gas of strings at thermal equilibrium  is  considered in Refs \cite{PatilBran,CosmoPheno}. In \cite{Estes:2011iw}, the case of the heterotic string compactified on a torus is analyzed at weak coupling. It is shown that at finite temperature, the points of enhanced gauge symmetry are minima of the free energy density, where all the internal moduli can be dynamically stabilized. There is no flat direction left (except for the dilaton) when the gauge group does not contain Abelian factors \cite{min}. In the S-dual picture in type I, one finds that the light vector multiplets responsible for the enhancement of the gauge group are either perturbative or D-strings  wrapped in the internal torus. In this case, the internal closed string moduli (Neuveu Schwarz-Neuveu Schwarz (NS-NS) and Ramond-Ramond (RR)) together with the open string Wilson lines are stabilized \cite{Estes:2011iw}. This indicates that  BPS states becoming massless at some point in moduli space should be treated on equal footing, wether they are perturbative or not.

In the present work, we use this fact to lift flat directions in the case of type II compactifications on CY threefolds, when finite temperature is switched on. Compared to the heterotic or type I strings on tori, the number of conserved supercharges present at zero temperature is half and the moduli space is by far more complicated. Due to $\N=2$ supersymmetry in four dimensions, it takes the form of a product $\M_V\times \M_H$ associated to Abelian vector multiplets and neutral hypermultiplets. Physically, these spaces realize the Coulomb and/or Higgs branches of Abelian and/or non-Abelian gauge theories  \cite{Strominger:1995cz, Katz,KlemmMyr, BKKM}. Due to the fact that the type II dilaton sits in the universal hypermultiplet, the metric on $\M_H$ admits corrections in string coupling. On the contrary, the metric on $\M_V$ is exact at tree level but is singular on loci, where 2-cycles in type IIA (3-cycles in type IIB) vanish \cite{Candelas:1989ug}. This fact is interpreted as the consequence of the existence of D2-branes (D3-branes) wrapping these cycles. They realize generically massive BPS states charged under the gauge group associated to the cycles and are integrated out, at the level of the low energy supergravity description. Therefore, when the cycles vanish and the BPS states become massless, the sigma-model metric on $\M_V$ develops a logarithmic divergence \cite{Strominger:1995cz}.\footnote{In the case of $\N=4$, the moduli space describes Coulomb branches only, which are not corrected by the string coupling and do not present infrared (IR) divergences, as follows from  the vanishing of the gauge beta functions.}

The aim of our work is to argue that at finite temperature, the moduli adjust so that a maximum number of 2-cycles, as well as 3-cycles vanish. To show this, we consider the low energy description of the models {\em without integrating out} the modes which become massless when the internal CY space is singular. Our approach can be summarized as follows.  By convention, we present it in type IIA compactified on a CY $M$, rather than in the equivalent mirror picture in type IIB.

\begin{description}
\item$(i)$ In the vicinity of a singular  point in $\M_V$, we identify the gauge group and charged matter arising from wrapped D2-branes on vanishing 2-cycles. Due to the fact the gauge theory is {\em non}-asymptotically free, a perturbative treatment in the low energy regime we are interested in is justified. We include these degrees of freedom as local fields in a tree level effective supergravity. The latter is insensitive to the temperature, since the Euclidean time circle can only be probed by loop corrections. The classical $\N=2$ supergravity is based on a product of special \Ka and quaternionic manifolds $\tilde\M_V\times \tilde\M_H$, whose metrics are unknown but satisfy constraints. First of all, they do not develop IR divergences and are thus regular. Second, they admit isometries we have to gauge in order to reproduce the gauge sector engineered geometrically.

\item $(ii)$ The gauging introduces a scalar potential we determine explicitly in the neighborhood under consideration. Its flat directions admit Coulomb and often Higgs branches. Moving from the Coulomb phase to the Higgs phase corresponds to an extremal transition from the original internal space $M$ to a topologically distinct CY space $M'$, where vanishing 2-cycles have been deformed into 3-cycles.
	
\item $(iii)$ In each branch, it is straightforward to determine from the potential the classical masses of the heavy states that belong to the gauge plus charged matter system. These masses depend on the moduli, which parameterize the flat directions associated to the Coulomb or Higgs phases.

\item $(iv)$ In the weak string coupling regime and at sufficiently low temperature/energy, the above masses are the only things needed to compute the one-loop correction to the effective supergravity. The result amounts to adding the one-loop effective potential $\F$ to the classical action evaluated in some tree level vacuum. One finds that {\em all flat directions in the Coulomb and Higgs branches of the geometrically engineered gauge theory are lifted}.

\item $(v)$ The one-loop action does not admit static solutions anymore. In other words, a cosmological evolution is induced by the thermal/quantum corrections. While the universe expands and the temperature drops, the moduli fields are attracted to the minimum of $\F$. The latter sits at the origin of the Coulomb and Higgs branches, where all tree level masses of the gauge plus matter system vanish. However, at one loop, all moduli masses are of order the temperature, while the gauge bosons remain massless. In fact, the only cosmological evolution with static moduli corresponds to the compactification on the singular configuration at the extremal transition between $M$ and $M'$.
\end{description}

The case of type II compactifications on CY spaces at finite temperature is useful to present mechanisms that may play a role in realistic descriptions of our world. However, since the temperature $T(t)$ converges to zero at late times, $\N=2$ supersymmetry is restored, which is in  obvious contradiction with zero temperature standard model phenomenology. In addition, the effective moduli masses arising from the free energy being of order $T(t)$, they  also vanish in this limit. Instead, realistic models would involve $\N=1$ supersymmetry for chiral matter to exist. Moreover, even at low temperature, $\N=1$ should be spontaneously broken and MSSM-like phenomenology be recovered at late times. Therefore, the analysis of the present work should be extended to type II backgrounds  involving orientifold projections, branes and internal fluxes \cite{geneCY} to describe $\N=1$ no-scale models at finite temperature. The latter are characterized by two independent supersymmetry breaking scales: The no-scale modulus $M$ and the temperature $T$. 

Toy $\N=1$ no-scale models at finite temperature have been studied in Scherk-Schwarz compactifications of the heterotic string on $T^6/(\Z_2 \times \Z_2)$  \cite{CriticalCosmo,stabmod,dSi}. Technically, the scale $M$ is introduced by imposing distinct  boundary conditions for bosons and fermions along  directions of the internal torus \cite{SS,KoR}. This is similar to the implementation of the temperature $T$ using the Euclidean time circle. As follows from the no-scale structure, it is relevant to split the moduli in two sets \cite{Ferrara:1994kg}:\\
\indent - Set 1 contains those which participate in the supersymmetry breaking, among which the no-scale modulus $M$ and the dilaton $\phi$ belong to.\footnote{The dilaton  participates in the supersymmetry breaking due to the fact that $M$ is dimensionfull and  acquires a dilaton dressing in Einstein frame.} \\
\indent - Set 2 contains the remaining moduli. \\

\noindent At weak coupling, the one-loop effective potential at finite temperature back-reacts on the flat background and induces a cosmological evolution. For suitable choices of supersymmetry breaking \cite{stabmod,CriticalCosmo}, the time-trajectory of the universe is attracted to a particular homogeneous and isotropic critical solution. The latter satisfies $T(t)\propto M(t)\propto e^{3\phi(t)}\propto 1/a(t)\propto 1/\sqrt{t}$, where $a$ is the scale factor and $t$ is cosmic time, while the other moduli are static. This evolution is radiation-like, in the sense that the coherent motion of $M(t)$ conspires with the free energy to imply the Hubble ``constant" to satisfy $H^2\propto T^4$, as if the universe were radiation dominated. Except for $M$ and $\phi$ which are running away, the moduli in set 1 have been attracted to minima of the thermal effective potential, where they have ``time-dependent masses" of order $M(t)^2/M_{Planck}$. For the moduli in set 2, a mechanism similar to that described in the present work applies. They acquire a mass of order $M(t)$ when they sit at points where the classical masses of bosons are vanishing. We stress that the fact the moduli masses are time-dependent implies that the energy density stored in the oscillations of these scalars around their minima is dominated by $T^4$ \cite{stabmod}.\footnote{When supersymmetry in four dimensions is broken by finite temperature only (no $M$), the energy of the oscillations  is of  order that of thermal radiation \cite{Estes:2011iw}.} Constant masses would instead lead to a density scaling as $T^3$, which would dominate over radiation and overclose the universe \cite{cosmomodprob}. 

The above solution is valid during an ``intermediate cosmological era", where $T(t)$ evolves between the Hagedorn temperature $T_H$ and a  scale $Q$ that should appear at  late times in realistic MSSM-like models, $T_H>T(t)> Q$. The upper bound expresses the fact that a Hagedorn transition may occur at earlier times.\footnote{Other scenarios can be imagined, where the intermediate era starts after a period of inflation followed by reheating.} The lower bound $Q$ is the ``IR renormalization group invariant transmutation scale" induced by the radiative corrections of the soft supersymmetry breaking terms at low energy\cite{AlvarezGaume:1983gj,NoscaleTSR}. For temperatures higher than $Q$, the renormalization group effects are negligible and the above described radiation-like evolution is legitimate. However, when $T(t)\le  Q$, they should induce the electroweak symmetry breaking and the stabilization of the supersymmetry breaking scale $M$ at a value of order $Q$. It is at this stage that all moduli in sets 1 and 2 acquire constant masses of order  $\langle M\rangle^2/M_{Planck}$ and $\langle M\rangle$, respectively.
In this scenario, the attraction to the critical solution where $M(t)\propto 1/\sqrt{t}$ followed by the stabilization of $M(t)$ around the scale $Q\sim 1$ TeV lead to a dynamical explanation of the hierarchy $M\ll M_{Planck}$. 

Returning to the present article, we present in details in Section \ref{coni} the program $(i)$--$(v)$ in the vicinity of a conifold locus. The gauge theory in this case is Abelian, with charged hypermultiplets \cite{Strominger:1995cz}. We show that the \Ka moduli of $M$ and complex structure moduli of $M'$ involved in the extremal  transition $M\leftrightarrow M'$ are attracted to this locus. A similar analysis is done in Section \ref{nonabel} in the neighborhood of points in the moduli space, where the internal CY $M$ develops a genus-$g$ curve of $A_{N-1}$ singularities (with $g\ge 1$) and can be deformed into another CY space $M^{\prime\prime}$ (when $g\ge 2$). This system describes an $SU(N)$ gauge theory coupled to $g$ hypermultiplets in the adjoint representation \cite{Katz}. In this case, \Ka and complex structure moduli of $M$, together with  complex structure moduli of $M^{\prime\prime}$, are attracted to the singular locus. In Section \ref{inters}, we argue that for any given internal CY space, we expect our approach to apply to all \Ka moduli and most of the complex structure moduli\footnote{To be specific, it is not clear to us if a complex structure controlling the size of vanishing 3-cycles that {\em cannot} be blown up to 2-cycles can be stabilized this way.}. However, the universal hypermultiplet scalars remain flat directions, at least in the weak coupling regime.
To illustrate our results, we consider the explicit example of  a type IIA compactification on a CY $M$ with Hodge numbers $(h_{11},h_{12})=(2,128)$. The moduli space $\M_V$ admits two codimension one loci, where $M$ develops either a node or a genus-2  curve of $A_1$ singularities. It follows that both \Ka moduli and some complex structure moduli can be stabilized at the intersection of these two loci. Given the fact that $M$ is a $K3$-fibration \cite{KlemmMayrK3}, the dual heterotic description \cite{DualConst}  on $K3\times T^2$   at finite temperature is known. It follows that the $T^2$ modulus $T_h$ and axio-dilaton $S_h$ are stabilized in the strong coupling regime. Section \ref{cl} summarizes our results and presents our perspectives.


\section{Stabilization at a conifold locus}
\label{coni}

In this Section, we consider the type II superstring compactified on $M$ or $M'$, two CY manifolds related by a conifold transition\footnote{In some particular case, there is no CY $M'$ in which the conifold $M$ can be deformed to.}. Our aim is to show that when finite temperature is switched on, the moduli involved in the extremal transition $M\leftrightarrow M'$ are lifted and attracted to the conifold locus, where they can be stabilized.
We choose to present our analysis in type IIA. Due to mirror symmetry, the type IIB picture can be derived by exchanging the roles of 2-cycles and 3-cycles.


\subsection{The geometrically engineered Abelian gauge theory}
\label{engicon}

At zero temperature, the compactification on $M$ yields an $\N=2$ theory in four dimensions. The massless spectrum contains in the gravitational multiplet the metric $g_{\mu\nu}$ and the graviphoton $A^0_\mu$ (from the RR 1-form). When $M$ is nonsingular, with Hodge numbers $h_{11}$ and $h_{12}$, there are $h_{11}$ \Ka deformations and $2h_{12}$ complex structure deformations of the CY metric. Combining these geometrical moduli with the reductions of the NS-NS 2-form and RR 3-form lead to the bosonic content of $h_{11}$ Abelian vector multiplets and $h_{12}$ neutral hypermultiplets.  Finally, the dilaton, the axion and the reduction of the 3-form on the unique $(3,0)$ and $(0,3)$ cycles realize the scalar content of the universal hypermultiplet. In total, the gauge group is $U(1)^{h_{11}+1}$. Moreover, due to the fact that $\N=2$ supersymmetry forbids couplings between vector multiplets and neutral hypermultiplets, the moduli space is a Cartesian product $\M_V\times \M_H$. The vector multiplet moduli space $\M_V$ of complex dimension $h_{11}$ is a special \Ka manifold, which is exact at tree level since the dilaton sits in a hypermultiplet. On the contrary, the hypermultiplet moduli space $\M_H$ of real dimension $4(h_{12}+1)$ is a quaternionic manifold, which admits perturbative and non-perturbative corrections. Classically, it is a product manifold, where an $SU(2,1)/U(2)$ factor is associated to the universal hypermultiplet.

By definition, along a conifold locus of codimension $S$ in $\M_V$, $S$  homology classes of 2-cycles are vanishing,  and  $R\ge S$ representative 2-cycles  in $M$ are shrinking to isolated points called nodes \cite{Candelas:1989ug}.
The metric of $\M_V$ appears in the  low energy  effective $\sigma$-model description of the vector multiplets. To account for the fact that this metric is singular along the conifold locus and cannot be cured by quantum corrections in string coupling,  a consistent picture has been proposed in Ref. \cite{Strominger:1995cz}. In this work, it is supposed that generically massive states charged under the $U(1)^S$ gauge factors have been integrated out, and become massless along the conifold locus. Consistently, the $\sigma$-model metric develops an IR  divergence\footnote{In gauge theory, this effect arises at one-loop in gauge coupling. Since the type II description of the $\N=2$ vector multiplet sector is exact in string coupling, the one-loop and non-perturbative corrections in gauge coupling are present at string tree level.}.  Since the gauge bosons arise from the RR 3-form, the charged states must be D2-branes\footnote{An electric-magnetic duality can always be used to work with purely electric D2-branes, without introducing magnetic D4-branes.}. To be massless when the homology classes vanish, the D2-branes must be BPS and wrapped on the shrinking 2-cycles. To reproduce the precise coefficient of the logarithmic divergence, the charged states must be hypermultiplets. The wrapped D2-branes being point-like from a four-dimensional point of view, they are extremal black hole hypermultiplets.

Because the local neighborhood of a node looks like a cone whose base is $S^2\times S^3$, the singular CY $M$ is called a conifold. When $R>S$, this configuration can be a passage to another smooth CY $M'$  obtained by deforming the shrinking 2-cycles into 3-cycles. This is the conifold transition, where the Hodge numbers $h'_{11}$ and $h'_{12}$ of $M'$ satisfy \cite{Candelas:1989ug}
\begin{align}
\label{sgr2}
h'_{11}= h_{11}-S\; , \qquad   h'_{12}= h_{12}+R-S\, .
\end{align}
Denoting by $\M'_V\times \M'_H$ the moduli space of $M'$, the extremal transition $M\leftrightarrow M'$ means $\M_V$ and $\M'_H$ are connected along the conifold locus. This geometrical picture matches perfectly  the physical interpretation of the system in terms of a $U(1)^S$ gauge theory coupled to $R$ hypermultiplets. $\M_V$ corresponds to the Coulomb branch: The scalars of the Abelian vector multiplets have nontrivial VEV's, while the charged states arise from massive non-perturbative D2-branes. $\M'_H$ corresponds to the Higgs branch: The Abelian vector multiplets combine with $S$ charged hypermultiplets to give $S$  massive long vector multiplets\footnote{It is not clear wether these massive multiplets are perturbative or non-perturbative.}. The remaining $R-S$ charged hypermultiplets are massless perturbative states, which condense \ie develop nontrivial VEV's along $\M_H'$.

Note that if the IR behavior of a $U(1)^S$ gauge theory is able to account for the singularity  of the \Ka metric on $\M_V$, this does not mean the theory remains Abelian in the ultra-violet. Actually, type II compactifications on CY spaces with moduli sitting in the vicinity of a conifold locus can engineer geometrically $\N=2$ asymptotically free non-Abelian gauge theories \cite{SWinString, X8}. In this case,  the non-Cartan gauge bosons are massive and our  description of the theory in terms of an Abelian gauge group is valid for low enough energies and temperatures.


\subsection{Tree level low energy description in gauged supergravity}
\label{sugracon}

To proceed, we determine the low energy description of the type IIA compactification on $M$ (and eventually $M'$, when a conifold transition is allowed), near a conifold configuration. At tree level in string coupling,  the result is insensitive to temperature effects, since genus-zero worldsheets cannot probe an Euclidean time circle.
To be consistent on both sides of the extremal transition, the $\N=2$ gauged supergravity we are looking for has to include all light and possibly massless degrees of freedom in the vicinity of  the conifold locus, wether they are realized perturbatively or non-perturbatively from the string point of view.

It is convenient to start our discussion from the perspective of the type IIA compactification on $M$. The effective action is constructed in two steps. First, we consider the ungauged $\N=2$ supergravity coupled to $h_{11}$ vector multiplets and $h_{12}+1+R$ hypermultiplets. The scalars  of the vector multiplets live on a special \Ka manifold $\tilde \M_V$, while those of the hypermultiplets span a quaternionic manifold $\tilde \M_H$. Both metrics $g_{I\bar J}$ and $h_{\Lambda\Sigma}$ on $\tilde \M_V$ and $\t \M_H$ are unknown, but satisfy properties we are going to use. In particular, they are regular even when $M$ is a conifold. By abuse of language, we will refer to the set of points in $\t \M_V\times \t\M_H$ corresponding to compactifications on conifold configurations of $M$ as the conifold locus.
There is a symplectic bundle over $\t \M_V$, whose holomorphic section admits electric and  magnetic components we denote by $X^0,\dots,X^{h_{11}}$ and $F_0,\dots,F_{h_{11}}$. The former can be used as homogeneous coordinates on $\t\M_V$. Thus, at a given point $P_0\in \t \M_V\times \t\M_H$ along the conifold locus, at least one of them, say $X^0$, does not vanish and can be set to 1 in a whole neighborhood of $P_0$. The remaining complex  components $X^I$ ($I=1,\dots,h_{11}$) are then the vector multiplet scalars and special coordinates on $\t \M_V$. In the vicinity of $P_0$, we also denote by $q^{\Lambda}$ ($\Lambda=1,\dots,4(h_{12}+1+R)$) a system of real coordinates parameterizing the hypermultiplet scalar manifold  $\t \M_H$.

In a second step, the charges of the hypermultiplets are introduced by gauging a $U(1)^S$ isometry group the quaternionic manifold $\t\M_H$ must satisfy. By convention, we label the vectors and scalar components of the gauged vector multiplets as $A^i_\mu$ and  $X^i$ ($i=1,\dots,S$), while the remaining $X^p$'s ($p=S+1,\dots,h_{11}$) denote the scalars of the ungauged ones. With these conventions, the tree level gauged supergravity action for the metric and  scalars takes the following form  \cite{Andrianopoli:1996cm},
\begin{align}
\label{PCL1}
S_{\rm tree}=\int d^4x\,\sqrt{-g}\left\{{\R\over 2}-g_{I\b J}\, \partial_\mu X^I\partial^\mu \b X^J - h_{\Lambda\Sigma}\nabla_{\!\!\mu} q^{\Lambda}\nabla^\mu q^{\Sigma} -\V\right\}\!,
\end{align}
where the covariant derivatives involve the non-trivial Killing vectors $k_i^\Lambda$,
\begin{align}
\nabla_{\!\!\mu} q^\Lambda=\partial_\mu q^\Lambda+A^i_\mu k_i^\Lambda,
\end{align}
and the scalar potential $\V$ is given by
\begin{align}
\label{PCL2}
\V=4 h_{\Lambda\Sigma} \,k^{\Lambda}_i \, \b k^{\Sigma}_j \,e^{\K} \b X^i  X^j+ g^{I \b J}f^i_I\, \b f^j_J\,\P^x_i\P^x_j -3\,e^\K  \b X^i  X^j\,\P^x_i\P^x_j.
\end{align}
In this expression, $\K$ is the \Ka potential associated to the metric $g_{I\b J}\equiv \partial_{X^I}\partial_{\b X^J}\K$,
\begin{align}
\label{Kp}
\K=-\ln\Big[i\Big(F_0-\b F_0+\b X^IF_I-X^I\b F_I\Big)\Big],
\end{align}
 and
\begin{align}
\label{sgr601a}
f^i_I=\Big(\partial_{X^I}+\demi \partial_{X^I} \K \Big)\big(e^{\demi\K}X^i\big),\qquad \b f^i_I=\Big(\partial_{\b X^I}+\demi \partial_{\b X^I} \K \Big)\big(e^{\demi\K}\b X^i\big).
\end{align}
Moreover, for each Killing vector, there is an $SU(2)$-triplet of  momentum maps $\P_i^x$, which are functions of $q^\Lambda$. They are related to the hyper-\Ka 2-forms $K^x$ on $\t \M_H$ by the relation
\begin{align}
\label{sgr603}
2\, k_i^\Lambda K^x_{\Lambda\Sigma}=\nabla^{SU(2)}_\Sigma \P^x_i\equiv\partial_{q^\Sigma} \P_i^x+\epsilon^{xyz}\omega^y_\Sigma \P^z_i,
\end{align}
where $\omega^x$ is the connection of the $SU(2)$-bundle. The fact that the Killing vectors $k_0^\Lambda$ and $k_p^\Lambda$ vanish identically implies the associated momentum maps are covariantly constant and thus trivial, $\P^x_0\equiv\P^x_p\equiv0$, as follows from the theorem recalled in Appendix C.

In a vacuum, the no-scale model has a vanishing potential, $\V=0$. To identify the conifold locus on $\t\M_V\times \t\M_H$, we use our knowledge of the geometrical realization of the gauge theory. When $M$ is a conifold, all multiplets in the action (\ref{PCL1}) must be massless. For the $q^\Lambda$'s to be massless, we see from the potential (\ref{PCL2}) that $\langle X^i\rangle=0$ is required, while  for the $X^i$'s to be massless, the Killing vectors and momentum maps must have zeros, $\langle k_i^\Lambda\rangle=\langle \P^x_i\rangle=0$. Thus, $P_0$ is fixed under the $U(1)^S$ isometries. In the remaining part of this Section, our aim is to expand the Lagrangian density in the action (\ref{PCL1}) around the point $P_0$.


\vskip .2cm
\noindent {\large \em Vector multiplet sector:}
We  start with the vector multiplet sector and denote the coordinates of $P_0$ in $\t\M_V$ as $(X^i_0=0,X^p_0)$. Smoothness of the  $\sigma$-model \Ka metric in (\ref{PCL1})  allows us to write\footnote{$\O(X-X_0)$ denotes without distinction holomorphic or antiholomorphic first order terms.}
\begin{align}
g_{I\b J}=g_{I\b J}^{(0)}+\O(X-X_0),
\end{align}
where $g_{I\b J}^{(0)}\equiv \left.g_{I\b J}\right\abs_{P_0}$. Moreover, we will also need the finite value $\K^{(0)}$ of the \Ka potential (\ref{Kp}) at $P_0$,
\begin{align}
\label{Ka0}
\K^{(0)}=-\ln\Big[i\Big(F^{(0)}_0-\b F^{(0)}_0+\b X_0^pF^{(0)}_p-X_0^p\b F^{(0)}_p\Big)\Big],
\end{align}
in terms of which the $f^i_I$'s defined in Eq. (\ref{sgr601a}) can be expressed as,
\begin{align}
\label{f}
f_I^i=e^{\half\K^{(0)}}\, \delta^i_I+\O(X-X_0),\qquad \b f_I^i=e^{\half\K^{(0)}}\, \delta^i_I+\O(X-X_0).
\end{align}


\vskip .2cm
\noindent {\large \em Hypermultiplet sector:}
The Taylor expansion in the hypermultiplet sector is more involved. In Appendix B, we show that on $\t\M_H$, there exists a new system of coordinates $c^{\A u}$ ($\A=1,\dots,R$; $u=1,2,3,4$), $q^{\alpha}$ ($\alpha=4R+1,\dots,4(h_{12}+1+R$)) such that $P_0$ is located at $(c^{\A u}=0,q^\alpha_0)$ and the complex structures $J^x$, the quaternionic metric and the hyper-\Ka forms at $P_0$ are:
\begin{align}
&\!\!\left.J^x\right|_{P_0}=-\eta^{xu}{}_v\left.\Big({\partial \over \partial c^{\ca u}}\otimes dc^{\ca v} \Big)\right\abs_{P_0}+\left.\Big(J^{x\alpha}{}_\beta\, {\partial \over \partial q^\alpha}\otimes dq^\beta \Big)\right\abs_{P_0}, \nonumber\\
&\!\!\left.h\right\abs_{P_0}= {1\over 2}\!\left.\big(dc^{\ca u}dc^{\ca u} \big)\right\abs_{P_0}+\left. \big(h_{\alpha\beta}\, dq^\alpha dq^\beta \big)\right\abs_{P_0}, \nonumber\\
&\!\!\left.K^x\right|_{P_0}={1\over 4} \eta^x_{uv}\left.\big(dc^{\ca u} \wedge dc^{\ca v} \big)\right\abs_{P_0}+{1\over 2}\!\left.\big(K^x_{\alpha\beta}\, dq^\alpha \wedge dq^\beta\big)\right\abs_{P_0},
\label{condJ}
\end{align}
where $\eta^{xu}{}_v$ are 't Hooft symbols defined in Appendix A.
In fact, the $(c^{\A1},\dots,c^{\A4})$'s are in the $\ca^{\rm th}$ hypermultiplets of charge $Q_i^\A$ under the $i^{\rm th}$ $U(1)$, while the remaining $q^\alpha$'s are the real components of the neutral ones\footnote{To make contact with the notations introduced in Appendix B, we define $c^{\A u}=\sqrt{2} q^{\A u}$ in order for the charged hypermultiplets to have canonically normalized kinetic terms. Moreover, we keep arbitrary the basis vector $\partial/\partial q^\alpha$ in the sub-tangent plane at $P_0$ associated to the neutral hypermultiplets.}. In order to write the kinetic terms and scalar potential in the vicinity of $P_0$, we need the expansions of the metric, hyper-\Ka forms and Killing vectors,\footnote{$\O(q-q_0)$ denotes terms of order $c^{\A u}$ or $(q^\alpha-q_0^\alpha)$.}
\begin{align}
\begin{array}{lll}
\dis h_{\A u,\B v}={1\over 2}\delta_{\A\B}\,  \delta_{uv}+\O(q-q_0), & h_{{\A u,\alpha}}=\O(q-q_0), & h_{\alpha\beta}=h^{(0)}_{\alpha\beta}+\O(q-q_0),\\
\dis K^x_{\A u,\B v}={1\over 2}\delta_{\A\B}\, \eta^x_{uv}+\O(q-q_0), & K^x_{{\A u,\alpha}}=\O(q-q_0), &K^x_{\alpha\beta}=K^{x(0)}_{\alpha\beta}+\O(q-q_0),\esp\\
\label{flat}k^{\A u}_i=Q_i^{\A u}\, t^u{}_v\, c^{\A u}+\O((q-q_0)^2)& \mbox{(no sum over $\A$),} & k^\alpha_i=\O((q-q_0)^2).
\end{array}
\end{align}
The first order terms of the Killing vectors involve $t^u{}_v$, a $U(1)$ generator acting on each hypermultiplet, and it is a matter of convention to choose $t^u{}_v\equiv-\b\eta^{3u}{}_v$ (see Appendices A and B). The charges $Q_i^\A$ are determined by the underlying CY geometry.
For this purpose, we define  $(\alpha^0,\dots,\alpha^{h_{11}})$ to be an homology basis of 2-cycles in $M$, among which $\alpha^i$ ($i=1,\dots,S$) vanish at the conifold locus. We also denote by $\gamma^{\ca}$ ($\ca=1,\dots,R$) the $R$ 2-cycles which shrink to nodes and expand $\gamma^{\ca}=n^{\ca}_i\alpha^i$. Then, the computation of the effective action on the world volume of a D2-brane wrapped on $\gamma^\A$ shows that $Q^{\ca}_i= n^{\ca}_i$  \cite{Strominger:1995cz}.

To determine the momentum maps $\P_i^x$, we first use the facts that they vanish at $P_0$ and that the left hand side of Eq. (\ref{sgr603}) is first order to conclude that the $\P_i^x$'s are actually second order. Next, we write Eq. (\ref{sgr603}) as
\begin{align}
{\partial\P^x_i\over \partial c^{\A u}}=Q_i^\A (\eta^xt)_{uv}c^{\A v}+\O((q-q_0)^2) \; \;\mbox{(no sum over $\A$)},\quad {\partial \P^x_i\over \partial q^\alpha} =\O((q-q_0)^2),
\end{align}
which we integrate to find
\begin{align}
\label{P}
\P^x_i=-{1\over 2}\, Q^\A_i\,c^{\A u}(\eta^xt)_{uv}c^{\A v}+\O((q-q_0)^3).
\end{align}


\vskip .2cm
\noindent {\large \em Effective action:}
From Eqs (\ref{f}), (\ref{flat}) and (\ref{P}), we find that in the potential $\V$ in Eq. (\ref{PCL2}), the two first terms which are positive are of order four in $(X-X_0)$ or $(q-q_0)$, while the last one, which is negative and characteristic of supergravity, is of order six and thus negligible around the point $P_0$. To write the potential $\V$ in an explicitly $SU(2)_R$-invariant form, we introduce the doublets
\begin{align}
\label{doublets}
\cC^\A=\left(\!\!\begin{array}{c} i(c^{\A 1}+ic^{\A 2})\\(c^{\A 3}+ic^{\A 4})^*\end{array}\!\!\right)
\end{align}
and obtain after some straightforward computation
\begin{align}
\label{PCL13}
\V= e^{\K^{(0)}}\Big(2 \, Q^{\ca}_i Q^{\ca}_j\, \b X^i X^j\, \mfs C^{\ca\dagger}\mfs C^{\ca}+ {1\over 4}\, g^{(0)i\b \jmath}\,D_i ^x  D_j^x \Big)+\cdots\;\;\;\where\;\;\; D_i^x\equiv Q_i^\A\, \mfs C^{\ca\dagger}\sigma^x\mfs C^{\ca}
\end{align}
are $SU(2)_R$-triplets of $D$-terms, $\sigma^x$ are the Pauli matrices and the ellipsis denote order five contributions in vector or hypermultiplet scalars.

In the end, close to a conifold configuration, the tree level effective action (\ref{PCL1}) associated to the type IIA superstring theory at finite temperature and compactified on either $M$ or $M'$ takes the final form,
\begin{align}
\label{Sconi}
S_{\rm tree}=\int d^4x \,\sqrt{-g}\,\bigg\{&{\R\over 2}-g^{(0)}_{I\b J}\, \partial_\mu X^I\partial^\mu \b X^J - {1\over 2}\,\nabla_{\!\!\mu} c^{\A u}\nabla^\mu c^{\A u} -h^{(0)}_{\alpha\beta}\,\partial_\mu q^\alpha\partial^\mu q^\beta\nonumber\\
&-e^{\K^{(0)}}\Big(2 \, Q^{\ca}_i Q^{\ca}_j\, \b X^i X^j\, \mfs C^{\ca\dagger}\mfs C^{\ca}+ {1\over 4}\, g^{(0)i\b \jmath}\,D_i ^x  D_j^x \Big)+\cdots\bigg\}.
\end{align}
It is interesting to note that the above action is that of the rigid $\N=2$ supersymmetric Abelian gauge theory with charged hypermultiplets and formally coupled to gravity.


\subsection{Lifting the Coulomb branch at one-loop}
\label{coniCoul}

The tree level scalar potential (\ref{PCL13}) valid around $P_0$ admits flat directions. We recall that due to the no-scale structure of the theory, these directions are insensitive to the scale of spontaneous symmetry breaking, here identified with the temperature. However, the picture is drastically modified once quantum corrections are taken into account. In fact, a moduli and temperature dependent correction to the classically vanishing vacuum energy is generated and, as we are going to see, lifts all classical flat directions associated to the Abelian gauge theory. In the following, our analysis is restricted to a weak string coupling regime, with quantum corrections computed at one-loop.

In the neighborhood of the point $P_0$ with coordinates $(X_0^i=0,X^p_0;c_0^{\A u}=0, q^\alpha_0)$ in $\t\M_V\times \t\M_H$, the set of vacua of the action (\ref{Sconi}) is a Cartesian product between:
\begin{itemize}

\item The space parameterized by the ``spectator moduli" $X^p$ and $q^\alpha$, which are only coupled gravitationally to the gauge theory. These scalars are coordinates along the conifold locus and reflect the arbitrariness in the choice of $P_0$ on it.

\item The space of configurations of the $X^i$'s and $c^{\A u}$'s, for which the semi-definite positive potential $\V$ vanishes. It is characterized by the constraints
\begin{align}
\label{vac}
\forall \A: ~~  X^i Q^\A_i\cC^\A=\left(\!\!\begin{array}{c}0\\0\end{array}\!\!\right)~(\mbox{no sum over $\A$})\qquad \and  \qquad \forall x, i: ~~D^x_i=0,
\end{align}
which admits Coulomb and Higgs branches.
\end{itemize}
The Coulomb branch of vacua corresponds to arbitrary values for the gauged  vector multiplets scalars and vanishing VEV's for those in the charged hypermultiplets:
\begin{align}
\label{coul}
\mbox{Coulomb branch : } \quad \Big\{\big(X^i \mbox{ arbitrary}, c^{\A u}=0\big)\Big\}\times\Big\{\big(X^p, q^\alpha\big) \mbox{ arbitrary}\Big\}.
\end{align}

To write the one-loop effective action at finite temperature in this branch, we evaluate the tree level part (\ref{Sconi}) in a background of the above form (\ref{coul}) and add the one-loop Coleman-Weinberg thermal effective potential $\F$,
\begin{align}
\label{Sconi1}
S_{\mbox{\scriptsize 1-loop}}=\int d^4x \,\sqrt{-g}\,\bigg\{{\R\over 2}-g^{(0)}_{I\b J}\, \partial_\mu X^I\partial^\mu \b X^J -h^{(0)}_{\alpha\beta}\,\partial_\mu q^\alpha\partial^\mu q^\beta-\F+\cdots\bigg\}.
\end{align}
The computation of  $\F$ is done in the Euclidean version of the theory, with time  compactified on a circle of perimeter equal to the inverse temperature. All degrees of freedom are imposed $(-1)^F$ boundary conditions along the temporal circle, where $F$ is the fermion number. For an arbitrary supersymmetric spectrum, the result is
\begin{align}
\label{freegene}
\F= - \int_0^{+\infty} {d\ell\over 2\ell}\, {1\over (2\pi \ell)^2}\,\sum_s e^{-{M_s^2 \ell\over2}}\, \sum_{\t m_0} e^{-{\t m_0^2\over 2\ell T^2}}\left(1-(-1)^{\t m_0}\right)\!,
\end{align}
where $T$ is the temperature and $M_s$ is the classical mass of each degenerate pair $s$ of boson and fermion. In this expression, $\ell$ is the proper time along the virtual loop wrapped $\t m_0$ times around the temporal circle and all dimension-full quantities are measured in Einstein frame. From a thermodynamical point of view, $\F$ is the free energy density associated to a perfect gas of bosons and fermions. In string compactifications where the supersymmetric spectrum at zero temperature is determined by a fully known conformal field theory, the expression (\ref{freegene}) can be derived from a vacuum-to-vacuum string amplitude in Euclidean time (and suitable $(-1)^F$ boundary conditions). As an example, this is the case for the heterotic string on $T^{10-D}$, which leads to a $D$-dimensional model whose exact spectrum  is known when $D\ge 6$ (so that no NS5-brane can wrap the internal torus) \cite{Estes:2011iw}. However in general, contributions of modes realized non-perturbatively from a string perspective cannot be captured by the CFT on the wordsheet. For instance, in the type I models S-dual to the above mentioned heterotic cases, the perturbative amplitude has contributions arising from fundamental open and closed strings and must be supplemented by additional terms associated to non-perturbative D1-branes  running into the virtual  loop. The role of BPS D1-branes wrapped on 1-cycles and becoming massless plays a  role in stabilizing the type I moduli \cite{Estes:2011iw} similar to what we are going to find here for wrapped D2-branes.

Returning to our present case of interest, the light masses in the vicinity of $P_0$ along the Coulomb branch can be found by inspection of  the bosonic action (\ref{Sconi}). The massless level includes the supergravity multiplet, the $I=1,\dots, h_{11}$ vector multiplets and the $\alpha=1,\dots ,h_{12}+1$ neutral hypermultiplets. Of course, this is not a surprise, since this is nothing but the perturbative massless spectrum arising from the type IIA  compactification on a smooth CY manifold $M$. The light squared masses of the $\A=1,\dots, R$ charged black hole hypermultiplets realized in the Coulomb phase as BPS D2-branes wrapped on vanishing 2-cycles are given by
\begin{align}
\label{masses}
M_\A^2=4\,e^{\K^{(0)}}\abs Q_i^\A X^i\abs^2+\cdots,
\end{align}
where the dots stand for higher order terms in scalar fields. The leading term is consistent with the standard mass formula of BPS black holes \cite{Strominger:1995cz,Ceresole:1995jg}. In particular, it acquires a dilaton dressing $e^{-\phi}$ once measured in string frame, as expected for D-brane masses\footnote{This makes a difference with the analysis of Refs \cite{Estes:2011iw,stabmod} in heterotic orbifold compactifications, where the additional massless modes are perturbative, with tree level masses measured in string frame independent of the dilaton.}. Close enough to $P_0$, all  other  masses $M_s$ are bounded from below and heavier than the charged black holes:  $M_s\ge M_{\rm min}>  M_\A$. Table \ref{table coni} summarizes the superfield content and associated scalar VEV's in the Coulomb branch\footnote{See \cite{Fayet:1978ig} for a general discussion about field contents following a Higgs mechanism in $\N=2$ supersymmetric gauge theories.}.
\begin{table}[h]
	\centering
	   \begin{tabular}{|c|c|c|c|c|c|c|c|}
		\cline{2-8}
		     \multicolumn{1}{c|}{} & \multicolumn{2}{c}{\text{Scalars acquiring VEV's}} & \multicolumn{5}{|c|}{\text{Superfields}} \\ \cline{2-8}
              \multicolumn{1}{c|}{}  & \raisebox{-0.1cm}{In vector} & \raisebox{-0.1cm}{In hyper-} & \multicolumn{3}{c|}{Vector multiplets} & \multicolumn{2}{c|}{Hypermultiplets}\\ \cline{4-8}
                \multicolumn{1}{c|}{}  &\raisebox{0.2cm}{multiplets}  & \raisebox{0.2cm}{multiplets} & $\substack{\text{Massless}\\ \text{(moduli)}}$ &$\substack{\text{Massive}\\ \text{short}}$ & $\substack{\text{Massive}\\ \text{long}}$ & $\substack{\text{Massless}\\ \text{(moduli)}}$ & \small{Massive} \\ \cline{1-8}  & & & & & & & \\
                $\begin{array}c \mbox{Coulomb} \\\mbox{phase}\end{array}$& $X^i$ & none & $S$ & 0 & $0$ & 0 & $R$  \\  & & & & & & & \\  \hline
         $\begin{array}c \mbox{Higgs} \\\mbox{phase}\end{array}$ & none & \multicolumn{1}{m{2.3cm}|}{$\mfs C^{\ca}$ mod. gauge orbits such that $D^x_i =0$} & 0 & 0 & $S$ & $R-S$ & 0
        \\ \hline
	   \end{tabular}
	\caption{\small Superfield contents in the Coulomb and Higgs branches (when $R>S$) associated to the $\N=2$ $U(1)^S$ gauge theory coupled to $R$ hypermultiplets, which is  encountered in the neighborhood of a conifold locus  in $\t \M_V\times \t\M_H$. The  scalars $X^p$ and $q^\alpha$ of the massless spectator vector multiplets and hypermultiplets  are not represented.}
	\label{table coni}
\end{table}

Integrating over $\ell$, the free energy density (\ref{freegene}) can be written as
\begin{align}
\label{PCL16}
\F=-T^4 \left\{ \Big(4+ 4h_{11}+4(h_{12}+1)\Big)G(0) +4\sum_{\A} G\Big({M_\A\over T}\Big)+\O\big(e^{-{M_{\rm min}\over T}}\big)\right\}\!,
\end{align}
where the function $G(x)$ is expressed in terms of a Bessel function of the second kind,
\begin{align}
\label{apG1}
G(x)=2\sum_{k\in \mathbb Z}\Big({x\over 2\pi|2k+1|}\Big)^2K_2\big(x|2k+1|\big),
\end{align}
and $G(0)$ is Stefan's constant for  radiation associated to a pair of massless boson and fermion,
\begin{align}
\label{apG3}
G(0)={\Gamma(2)\over \pi^2}\sum_{k\in \mathbb Z} {1\over |2k+1|^4 }={\pi^2\over 48}.
\end{align}
Moreover, the first factor 4 in Eq. (\ref{PCL16}) corresponds to the $2+2$ on shell degrees of freedom of the graviton and graviphoton, while the other factors 4 count the number of bosonic degrees of freedom in vector multiplets and hypermultiplets.
In fact, for positive $x\ge 0$, the function $G(x)$ is maximum at the origin and decreases exponentially,
\begin{align}
\label{apG2}
G(x) =G(0)-{x^2\over 16}+ {\cal O}(x^4) \text{ when } x\ll 1, \quad G(x)\sim \Big({x\over 2\pi}\Big)^{3\over 2}e^{-x} \text{ when } x\gg1.
\end{align}
As a result, all contributions $G(M_s/T)$ with $M_s\ge M_{\rm min}$ are exponentially suppressed, provided the temperature is low enough, $T<M_{\rm min}$, as indicated in Eq. (\ref{PCL16}).

Since the free energy density $\F$ depends on the black hole hypermultiplet masses given in Eq. (\ref{masses}), it acts as a non-trivial potential for the scalars $X^i$. The behavior of $G(x)$ at $x=0$ implies $\F$ is minimum when all $M_{\A}$'s vanish \ie $\forall \A,\;  Q^\A_ i X^i=0$. Due to the fact that the matrix $Q^\A_i$ is of rank $S$,\footnote{Otherwise, some of the $R$ vanishing 2-cycles would be linear combinations of the others and would not give independent degrees of freedom once wrapped with D2-branes.} this can only happen at the conifold locus, $X^i=0$. In other words, all classically flat directions $X^i$ of the Coulomb branch are lifted, while the spectator scalars $X^p$ and $q^\alpha$ parameterizing  the conifold locus remain moduli. To find the one-loop masses $M_{i\mbox{\scriptsize 1-loop}}$ of the fields $X^i$ at their minimum, we consider the squared mass matrix
\begin{align}
{\Lambda^{\b I}}_{\b J}=g^{(0)\b I K}\!\left. {\partial^2\F\over \partial X^K\partial \bar X^J}\right\abs_{X^i=0}= {T^2\over16}\left.g^{(0)\b I K}\, 4\sum_{\A}{\partial^2 M^2_{\A}\over \partial X^K\partial \b X^J}\right\abs_{X^i=0},
\end{align}
which satisfies
\begin{align}
\Lambda^{\b I}{}_{\b \jmath}=T^2e^{\K^{(0)}}g^{(0) \b I k}\, Q^\A_kQ^\A_j\, , \qquad \Lambda^{\b I}{}_{\b p}=0 .
\end{align}
$\Lambda$ is diagonalizable, with $S$ strictly positive eigenvalues $M^2_{i\mbox{\scriptsize 1-loop}}$ and $h_{11}-S$ vanishing ones, so that the trace leads to
\begin{align}
\label{mass vec}
\sum_i M^2_{i\mbox{\scriptsize 1-loop}}=T^2e^{\K^{(0)}}g^{(0) \b \jmath k}\, Q^\A_kQ^\A_j.
\end{align}
Thus, the  scalars of the $U(1)^S$ vector multiplets acquire one-loop masses of order the temperature scale, while the gauge bosons remain massless and the full Abelian gauge theory $U(1)^{h_{11}+1}$ is unbroken. If the fact that vector multiplet scalars are getting masses is certainly a good thing, one may wonder if this result is not spoiled by the appearance of additional massless black hole hypermultiplets precisely at the minimum. In fact, the tree level masses of the scalars $c^{\A u}$ are vanishing at the conifold locus, but are acquiring non-trivial corrections of order $T$ at the one-loop level, as will be seen in the next Section in the study of the Higgs branch.

For a homogeneous and isotropic universe, the one-loop energy density and pressure found by varying the action (\ref{Sconi1}) are consistent with thermodynamics \cite{CriticalCosmo},
\begin{align}
\label{rhoP}
\rho=\F-T\, \frac{\partial \F}{\partial T} \; , \qquad \qquad P=-\F.
\end{align}
When $T/M_{\rm min}$ is low enough, they satisfy at the conifold locus the state equation for radiation, $\rho=3P$, as follows from Eq. (\ref{PCL16}). As a result, a particular solution to the equations of motion for the scale factor $a$, the temperature $T$ and the scalars is
\begin{align}
\label{solpart}
a(t)\propto {1\over T(t)}\propto \sqrt{t}\; , \quad X^i\equiv c^{\A u}\equiv 0\; , \quad X^p, q^\alpha \,\mbox{constant},
\end{align}
where $t$ is the cosmological time.
This evolution describes an expanding universe filled with radiation and static scalars. Consistently, the  temperature drops and guarantees  $T\ll M_{\rm min}$ is satisfied at late times. More general solutions consistent with homogeneity and isotropy exist, as follows from the general analysis of Ref. \cite{Estes:2011iw}. They are characterized by damped oscillations of the fields $X^i$, which converge to their minimum at $X^i=0$. The energy density stored in their oscillations  and in the motion of the spectator moduli $X^p$ and $q^\alpha$ scales as $T^4$. Thus, they do not dominate over radiation and the cosmological moduli problem \cite{cosmomodprob} is avoided. This follows from the fact that the masses of the scalars $X^i$ are actually proportional to the temperature and thus time-dependent \cite{Estes:2011iw}. To put a halt to the fall of these masses and obtain realistic models at low temperatures, one may follow the lines of Ref. \cite{stabmod}. At zero temperature, interesting models should be characterized by  $\N=1$ supersymmetry spontaneously broken at a scale $M$. Once finite temperature is taken into account, one finds the evolutions of the one-loop masses of the moduli, the temperature, the supersymmetry breaking scale $M$ and the inverse of the scale factor $1/a(t)$ are attracted to a particular trajectory, where they are proportional. When the temperature reaches the electroweak symmetry breaking scale $M_{ew}$, the moduli masses and $M$ are expected to be stabilized around $M_{ew}$, while $T$ keeps on decreasing.


\subsection{Lifting the Higgs branch at one-loop}
\label{coniHiggs}

Our aim in this Section is to complete the analysis of the conifold locus by showing that the Higgs branch of the $U(1)^S$ gauge theory is lifted by the one-loop thermal effective potential. In this phase, the doublets $\cC^\A$ are such that the D-terms in Eq. (\ref{vac}) vanish, while the $U(1)^S$ vector multiplet scalars have trivial VEV's,
\begin{align}
\label{higgs}
\mbox{Higgs branch : } \quad \Big\{\big(X^i =0, \cC^\A\mbox{ such that } D_i^x=0\big)\Big\}\times\Big\{\big(X^p, q^\alpha\big) \mbox{ arbitrary}\Big\}.
\end{align}
The $3S$ D-term constraints leave $4R-3S$ flat directions among the charged scalars $c^{\A u}$'s, along which the $U(1)^S$ gauge group is Higgsed. $S$ of the $4R-3S$ directions are orbits of the residual global $U(1)^S$ symmetry corresponding to physically equivalent vacua. We therefore introduce $S$ gauge fixing conditions reflecting the fact that $S$ would-be-Goldstone bosons are eaten by the massive vector fields. We are left with $4(R-S)$ flat directions, which can be arranged in $R-S$ massless neutral hypermultiplets. Clearly, for the Higgs branch to exist, $R>S$ is required. Moreover, the $S$ Higgsed vector multiplets become massive and long by combining with the remaining $S$ hypermultiplets. The superfield content and VEV's  in the Higgs branch can be found in Table \ref{table coni}.
Thus, besides the supergravity multiplet, the massless spectrum includes $h_{11}-S$ vector multiplets and $h_{12}+R-S+1$ neutral hypermultiplets, corresponding exactly to the type IIA compactification on the smooth CY manifold $M'$, with Hodge numbers $h'_{11}$ and $h'_{12}$ given in Eq. (\ref{sgr2}).

To describe the one-loop effective action in the Higgs branch, it is convenient to parameterize the D-term flat directions with some coordinates $\xi^m$ ($m=1,\dots,4(R-S)$) satisfying $Q^{\ca}_i\, \mfs C^{\ca\dagger}(\xi)\, \sigma^x\, \mfs C^{\ca}(\xi)=0$ and such that the Jacobian matrix $\Big({\partial c^{\ca u}\over \partial \xi^m}\Big)$ is of rank $4(R-S)$. We denote by $\xi^m_0$ the origin of the Higgs branch \ie the conifold locus. In a neighborhood of $P_0$, the one-loop effective action valid in the Higgs branch takes the form,
\begin{align}
\label{PCL17}
S_{\mbox{\scriptsize 1-loop}}=\int d^4x \, \sqrt{-g}\,\bigg\{{\R\over 2}  -g^{(0)}_{p\b q}\,\partial X^p\partial \b{X}^q -h_{mn}^{(0)} \, \partial \xi^m \partial \xi^n-h^{(0)}_{\alpha\beta}\, \partial q^{\alpha} \partial q^{\beta} -\F\bigg\},
\end{align}
where we have defined
\begin{align}
 h_{mn}^{(0)} ={1\over 2}\!\left.{\partial c^{\A u} \over \partial \xi^m}\right\abs_{\xi_0}\left.{\partial c^{\A u} \over \partial \xi^n}\right\abs_{\xi_0},
 \end{align}
and the free energy density $\cf$ is
\begin{align}
\label{PCL18}
\F=-T^4 \left\{ \Big(4+ 4h'_{11}+4(h'_{12}+1)\Big)G(0) +8\sum_{i} G\Big({M_{i}\over T}\Big)+\O\big(e^{-{M_{\rm min}\over T}}\big)\right\}\!.
\end{align}
The factor 8 in the above expression counts the number of boson/fermion pairs in the long vector multiplets of tree level mass $M_i$. The $\O\big(e^{-{M_{\rm min}\over T}})$ term includes all contributions of the modes, whose masses cannot vanish in the neighborhood we are considering and thus admit a lower bound $M_{\rm min}>M_i$. For $T<M_{\rm min}$, these contributions are exponentially suppressed.

To proceed, we determine the sum of the squared masses of the long vector multiplets. This can be derived from the first term of the tree level potential $\V$ in Eq. (\ref{PCL13}), when the hypermultiplet scalars $c^{\A u}$ condense along the D-flat directions. At second order in scalar fields, the matrix of squared masses of the $X^I$'s is
\begin{align}
\Delta^{\b I}{}_{\b \jmath}=2e^{\K^{(0)}}g^{(0) \b I k}\, Q^\A_kQ^\A_j\, c^{\A u}c^{\A u}+\cdots, \qquad \Delta^{\b I}{}_{\b p}=\cdots ,
\end{align}
whose trace gives
\begin{align}
\label{mass2}
\sum_i M^2_{i}=2e^{\K^{(0)}}g^{(0) \b \jmath  k}\, Q^\A_kQ^\A_j \, c^{\A u}c^{\A u}+\cdots.
\end{align}
Alternatively, the second term of $\V$ in Eq. (\ref{PCL13})  can be used to compute the sum of the squared masses in the $c^{\A u}$-sector,  when they condense. Consistently, the result in Eq. (\ref{mass2}) is recovered.

In the vicinity of $P_0$, the thermal effective potential $\F$ is minimal when all masses $M_i$ vanish. This occurs only at the origin of the Higgs branch, $c^{\A u}=0$,  \ie along the conifold locus.   Thus, all classically flat directions $\xi^m$ are lifted, while the neutral components $q^\alpha$ remain moduli.  The one-loop squared masses $M^2_{m\mbox{\scriptsize 1-loop}}>0$ of the fields $\xi^m$ at their minimum $\xi=\xi_0$ are determined by the matrix
\begin{eqnarray}
\!\!\!\!\!\!\!\!\!\!\!\!\!\!\!\!\!\!{\Lambda^{\prime m}}_n\!\!\!&=&\!\!\!{1\over 2}h^{(0)ml}\left.{\partial^2\F\over \partial\xi^l\partial\xi^n}\right\abs_{\xi_0}={T^2\over 16}\left. {1\over 2}h^{(0)ml} \, 8\sum_i{\partial^2 M^2_i\over \partial\xi^l\partial\xi^n}\right\abs_{\xi_0}\nonumber\\
\!\!\!&=& \!\!\!T^2e^{\K^{(0)}}g^{(0) \b \imath j}\, Q^\A_jQ^\A_i\, h^{(0)ml}\left.{\partial c^{\A u} \over \partial \xi^l}\right\abs_{\xi_0}\left.{\partial c^{\A u} \over \partial \xi^n}\right\abs_{\xi_0},
\end{eqnarray}
where  we have used the fact that $c^{\A u}\abs_{\xi_0}=0$ to obtain the last equality. The eigenvalues of $\Lambda'$ are the desired squared masses we are looking for. Taking the trace, they satisfy
\begin{align}
\label{mass hyp}
\sum_m M^2_{m\mbox{\scriptsize 1-loop}}=T^2e^{\K^{(0)}}g^{(0) \b \imath j}\, Q^\A_jQ^\A_i\, h^{(0)nl}\left.{\partial c^{\A u} \over \partial \xi^l}\right\abs_{\xi_0}\left.{\partial c^{\A u} \over \partial \xi^n}\right\abs_{\xi_0}.
\end{align}
Thus, the $\xi^m$'s have acquired a mass of order the temperature scale. Due to the arbitrariness in the choice of parametrization $\xi^m$ of the D-flat directions, we conclude that all charged black hole hypermultiplets scalars $c^{\A u}$ have a mass of order $T$. Combining this fact with the result of Eq. (\ref{mass vec}), we see that at the one-loop level, all scalars involved in the $U(1)^S$ gauge theory coupled to $R$ black hole hypermultiplets are acquiring masses at the conifold locus.

The particular homogeneous and isotropic evolution (\ref{solpart}) found in the study of the Coulomb branch can now be seen as a particular case of another set of cosmological solutions, where the scalars $\xi^m$ are oscillating with damping, as can be shown along the lines of Ref. \cite{Estes:2011iw}. The energy stored in the oscillations and in the motion of the spectator moduli $q^\alpha$ scales as $T^4$ and do not dominate over radiation.


\section{Stabilization at a  non-Abelian gauge symmetry locus}
\label{nonabel}

In this Section, our aim is to show how the moduli involved in extremal transitions realizing non-Abelian gauge theories can be stabilized, once finite temperature effects are taken into account.  We specialize on the case of a geometric description of an $SU(N)$ gauge theory coupled to $g\ge 1$ hypermultiplets in the adjoint representation.


\subsection{The geometrically engineered non-Abelian gauge theory}
\label{engicon2}

Our starting point is the type IIA theory compactified on a CY manifold $M$, which can develop a genus-$g$ curve $\C$ of $A_{N-1}$ singularities \cite{Katz,KlemmMyr}. Among the $h_{11}$ homology 2-cycles, $N-1$ are realized by 2-spheres $\Gamma_i$ ($i=1,\dots,N-1$)  in $M$, with intersection matrix corresponding to the Dynkin diagram of $A_{N-1}$, and with volume shrinking to zero when we sit on a codimension $N-1$ locus in the complexified \Ka moduli space $\M_V$. All  connected 2-cycles built out of the $\Gamma_i$'s are of the form $\Gamma_{ij}=\Gamma_i\cup\cdots\cup\Gamma_j$, for $1\le i\le j\le N-1$, and can be wrapped by BPS D2-branes or anti-D2-branes (obtained by reversing the orientations). The former (latter) are associated to the $(N^2-N)/2$ positive (negative) roots of $A_{N-1}$, while the perturbative spectrum provides the remaining massless multiplets in the Cartan subalgebra. In the large volume limit of the curve $\C$, the model leads to an $\N=2$ theory in six dimensions describing an $SU(N)$ gauge theory \cite{aspin}. Thus, one can think about the four dimensional case as arising from an additional compactification on the curve $\C$ of genus $g$, which breaks further half of the supersymmetries. The result is an $\N=2$ $SU(N)$ gauge theory coupled to $g$ hypermultiplets in the adjoint representation \cite{Katz}.

In the following, we restrict our attention to the cases where $g\ge1$. Actually, when $g=0$, the pure $SU(N)$ gauge theory is asymptotically free and, as already mentioned at the end of Section \ref{engicon}, is Abelian in the IR, with gauge group $U(1)^{N-1}$. Thus, this situation is expected to be dual to a particular example of the conifold case we have already studied, for $R=S=N-1$. For $g=1$, the vector and hypermultiplet in the adjoint representation  combine into an  $\N=4$ $SU(N)$ gauge sector. This case is conformal and has already been considered in Ref. \cite{Estes:2011iw}, yielding to an attraction of the moduli at the origin of the Coulomb branch, thus restoring the full non-Abelian symmetry. On the contrary, new physics is encountered for $g\ge 2$, since the $SU(N)$ gauge theory is non-asymptotically free and moreover admits  Coulomb and Higgs branches.

The above phases are realized geometrically by compactifying on the original manifold $M$ or on a distinct CY $M^\dprime$ related to $M$ by extremal transition, $M\leftrightarrow M^\dprime$. It is instructive to recover the Hodge numbers $h^\dprime_{11}$ and $h_{12}^\dprime$ of $M^\dprime$ derived by deforming the vanishing 2-cycles  into finite volume 3-cycles  \cite{Katz} from the gauge theory point of view. On a smooth CY $M$, which corresponds to a generic point in the Coulomb branch\footnote{\label{SU}When some non-trivial  VEV's in the Cartan subalgebra coincide, $SU(N)$ is broken to a product of non-Abelian $SU$ groups and $U(1)$ factors, with total rank equal to $N-1$. Geometrically, this corresponds to singular configurations of $M$, where some (but not all) of the $N-1$ 2-spheres are vanishing. These circumstances are encountered in loci in the moduli space that admit the full enhanced $SU(N)$ enhanced symmetry locus as a submanifold.}, the $SU(N)$ gauge group is spontaneously broken to $U(1)^{N-1}$, beside a remaining ``spectator" $U(1)^n$ factor, where $n=h_{11}-(N-1)+1$. The $h_{12}+1$ massless hypermultiplets include the $g(N-1)$ perturbative Cartan components of the $g$ adjoint matter representations, together with $m$ ``spectator" hypermultiplets, $m=h_{12}-g(N-1)+1$. The left-over $N^2-N$ non-Cartan vector and $g(N^2-N)$ matter multiplets are massive. On the other hand, for the compactification on $M^\dprime$ to reproduce the spectrum in the Higgs branch, one must have $h^\dprime_{11}+1= n$ Abelian vector multiplets, and $h^\dprime_{12}+1=(g-1)(N^2-1)+m$ massless hypermultiplets. The remaining $N^2-1$ matter multiplets combine with the $SU(N)$ vector multiplets into long massive vector multiplets. As a result, one obtains \cite{Katz}
\begin{align}
\label{h''}
h^\dprime_{11}= h_{11}-(N-1)\; , \qquad   h^\dprime_{12}= h_{12}+(g-1)(N^2-1)-g(N-1)\, .
\end{align}


\subsection{Tree level low energy description in gauged supergravity}
\label{sugranonA}

At finite temperature, the one-loop low energy effective action of the type IIA theory compactified on either $M$ or $M^\dprime$ (when $g\ge 2$) can be decomposed in two parts. The classical one, which is independent of $T$, and the one-loop Coleman-Weinberg thermal effective potential evaluated in some classical background. In this Section, we focus on the tree level action, which can be described by an $\N=2$ gauged supergravity. The latter, to be valid when $M$ (or $M^\dprime$) develops the curve of $A_{N-1}$ singularitites, has to include explicitly the whole set of light degrees of freedom, including the $SU(N)$ gauge sector coupled to $g$ hypermultiplets in the adjoint.

Due to the relation (\ref{h''}), we can organize our discussion in terms of the Hodge numbers of $M$. The ungauged supergravity we start with contains $h_{11}+N^2-N$ vector multiplets and $h_{12}+1+g(N^2-N)$ hypermultiplets, in order to take into account the non-Cartan generators, which arise from solitonic D2-branes when the compactification is on $M$. The scalar fields span a product manifold $\t \M_V\times \t \M_H$, with special \Ka and quaternionic metrics $g_{I\bar J}$ and $h_{\Lambda\Sigma}$. As in the case of the conifold locus, we refer to the set of points in $\t \M_V\times \t \M_H$ associated to compactifications developing the curve of $A_{N-1}$ singularities as the non-Abelian locus. In the symplectic bundle  $(X^0,X^I;F_0,F_I)$ ($I=1,\dots,h_{11}+N^2-N$), the electric components are homogeneous coordinates of $\t\M_V$. Therefore, in the neighborhood of a given point $P_0$ on the non-Abelian locus, we can set one of the $X$-entries to 1, say $X^0$, and work with special coordinates $X^I$.   Furthermore, we choose a chart $q^{\Lambda}$ $\big(\Lambda=1,\dots,4(h_{12}+1+g(N^2-N))\big)$ of real coordinates on the hypermultiplet manifold $\t \M_H$, whose properties will be specified shortly.

By construction, the metrics $g_{I\b J}$ and $h_{\Lambda\Sigma}$ are non-singular and admit a subgroup $SU(N)$ of isometries we now gauge. We choose the vectors and scalars of the gauged vector multiplets to be labeled as $A^a_\mu$ and  $X^a$ ($a=1,\dots,N^2-1$) and denote the left-over  ``spectator" scalars as $X^p$ ($p=N^2,\dots,h_{11}+N^2-N$). Restricting for simplicity to the metric and scalar degrees of freedom, the $\N=2$ tree level gauged supergravity action is
\begin{align}
\label{Scla}
S_{\rm tree}=\int d^4x\,\sqrt{-g}\left\{{\R\over 2}-g_{I\b J}\nabla_{\!\!\mu} X^I\nabla^\mu \b X^J - h_{\Lambda\Sigma}\nabla_{\!\!\mu} q^{\Lambda}\nabla^\mu q^{\Sigma} -\V\right\}\!,
\end{align}
where the covariant derivatives are expressed in terms of the non-trivial Killing vectors $k_a^I$ and $k_a^\Lambda$ 	acting on $\t\M_V$ and $\t\M_H$,
\begin{align}
\nabla_{\!\!\mu} X^I=\partial_\mu X^I+A^a_\mu k_a^I\, , \qquad\nabla_{\!\!\mu} \b X^I=\partial_\mu \b X^I+A^a_\mu \b k_a^I\, , \qquad \nabla_{\!\!\mu} q^\Lambda=\partial_\mu q^\Lambda+A^a_\mu k_a^\Lambda.
\end{align}
The scalar potential $\V$ takes the form \cite{Andrianopoli:1996cm}
\begin{align}
\label{Vcla}
\V=\left( g_{I\b J}\, k_a^I\b k_b^J+4 h_{\Lambda\Sigma} \,k^{\Lambda}_a \, k^{\Sigma}_b \right) e^{\K} \b X^a  X^b+ g^{I \b J}f^a_I\,\b f^b_J\,\P^x_a\P^x_b -3\,e^\K  \b X^a  X^b\,\P^x_a\P^x_b,
\end{align}
where the \Ka potential $\K$ is defined as in Eq. (\ref{Kp}) and
\begin{align}
\label{sgr601}
f^a_I=\Big(\partial_{X^I}+\demi \partial_{X^I} \K \Big)\big(e^{\demi\K}X^a\big),\qquad \b f^a_I=\Big(\partial_{\b X^I}+\demi \partial_{\b X^I} \K \Big)\big(e^{\demi\K}\b X^a\big).
\end{align}
The triplets of momentum maps appearing in Eq. (\ref{Vcla}) are associated to the non-trivial Killing vectors acting on the quaternionic manifold $\t\M_H$, and are related to the hyper-\Ka 2-forms $K^x$,
\begin{align}
\label{momna}
2\, k_a^\Lambda K^x_{\Lambda\Sigma}=\nabla^{SU(2)}_\Sigma \P^x_a\equiv\partial_{q^\Sigma} \P_a^x +\epsilon^{xyz}\omega^y_\Sigma \P^z_a.
\end{align}
Utilizing Appendix C, the momentum maps  $\P^x_0\equiv\P^x_p$ vanish identically, since the associated Killing vectors $k_0^\Lambda$ and $k_p^\Lambda$ are trivial.

The vacua satisfy $\V=0$, and in particular the non-Abelian locus in $\t \M_V\times \t\M_H$, where all vector multiplets and hypermultiplets we have introduced are massless. This is the case for the $q^\Lambda$'s if $\langle X^a\rangle=0$, and for the $X^a$'s if the Killing vectors and momentum maps vanish, $\langle k_a^\Lambda\rangle=\langle \P^x_a\rangle=0$. Thus, the point $P_0$ and actually the non-Abelian locus on which it sits are fixed by the isometries. In the following, our aim is to derive the Taylor expansion of the Lagrangian density of the action (\ref{Scla}) at $P_0$.


\vskip .2cm
\noindent {\large \em Vector multiplet sector:}
Let us denote the coordinates of $P_0$ in $\t\M_V$ as $(X_0^a=0,X^p_0)$ and begin our discussion with the vector multiplet sector. The infinitesimal isometry action on the scalars, $\delta X^I=\epsilon^ak_a^I$, is generated by Killing vectors satisfying the $su(N)$ algebra,
\begin{align}
\big[ k_a^I\partial_{X^I}\;,\;  k_b^J\partial_{X^J}\big]=f^{abc} \; k_c^I\partial_{X^I},
\end{align}
where $f^{abc}$ are structure constants. Since the $X^a$'s are in the adjoint representation and the Killing vectors are vanishing at $P_0$, we conclude that at lowest order,
\begin{align}
\label{knav}
k_a^b=f^{abc}X^c+\O((X-X_0)^2),\qquad k_a^p= \O((X-X_0)^2).
\end{align}
Utilizing Killing's equation, the above expressions can be used to constraint the zero$^{\rm th}$ order of the \Ka metric,
\begin{align}
\forall I,\b J, a:\quad g_{I\bar K}\, D_{\b J}\b k^K_a+g_{K\b J}\, D_Ik^K_a=0\quad \Longrightarrow\quad
g^{(0)}_{I\b b}\, {\partial (f^{abc}\b X^c)\over \partial \b X^J}+g^{(0)}_{b\b J}\, {\partial (f^{abc}X^c)\over \partial X^I}=0,
\end{align}
where $D$ is the covariant derivative on the complex manifold $\t\M_V$. Taking $(I,\b J)=(d,\b e)$ leads to $[g^{(0)},T^a]=0$, where $(T^a)_{bc}=-if^{abc}$ are the $SU(N)$ generators. Therefore, $g^{(0)}_{a\b b}$ is proportional to the identity matrix. On the other hand, the choice $(I,\b J)=(d,\b p)$ yields $g^{(0)}_{e\b p}=0$. Altogether, we conclude that there exists a constant $l^2>0$ such that
\begin{align}
\label{gna}
g_{a\b b}=l^2\delta_{a\b b}+\O(X-X_0), \qquad g_{a\b p}=\O(X-X_0),\qquad g_{p\b q}=g^{(0)}_{p\b q}+\O(X-X_0).
\end{align}
Finally, the \Ka potential at $P_0$ takes the form given in Eq. (\ref{Ka0}), in term of which we have
\begin{align}
\label{fna}
f_I^a=e^{\half\K^{(0)}}\, \delta^a_I+\O(X-X_0),\qquad \b f_I^a=e^{\half\K^{(0)}}\, \delta^a_I+\O(X-X_0).
\end{align}


\vskip .2cm
\noindent {\large \em Hypermultiplet sector:}
The matter sector necessitates more technical manipulations. Again, we start with  the Killing vectors, whose action on $\t\M_H$ must satisfy
\begin{align}
\big[ k_a^\Lambda\partial_{q^\Lambda}\;,\;  k_b^\Sigma\partial_{q^\Sigma}\big]=f^{abc} \; k_c^\Lambda\partial_{q^\Lambda}.
\end{align}
Since the geometry of the compactification is telling us that this algebra is realized by $4g$ adjoint representations, we know we can single out a coordinate system  $(q^{a\lambda},q^\alpha)$, where $a=1,\dots,N^2-1$ and $\lambda=1,\dots,4g$, and $\alpha=4g(N^2-1)+1,\dots, 4(h_{12}+1+g(N^2-N))$.   Physically, the $4g$ adjoint representations are the components of the $g$ hypermultiplets that are transforming under $SU(N)$, while the remaining $q^\alpha$'s are singlets of $SU(N)$ and therefore referred to as the components of the ``spectator" hypermultiplets. Denoting the coordinates of $P_0$ in $\t\M_H$ as $(q_0^{a\lambda},q_0^\alpha)$, where the Killing vectors are vanishing, the latter can be written at lowest order as
\begin{align}
\label{k0}
k_a^{b\lambda}=f^{abc}(q^{c\lambda}-q^{c\lambda}_0)+\O((q-q_0)^2),\qquad k_a^\alpha= \O((q-q_0)^2).
\end{align}

To write down the form of the quaternionic metric at $P_0$, we make use of Killing's equation, where we denote interchangeably the coordinate system as $q^\Lambda$ or $(q^{a\lambda},q^\alpha)$,
\begin{align}
\forall \Lambda,\Sigma, a:\quad h_{\Lambda\Xi}\, D_\Sigma k^\Xi_a+h_{\Sigma\Xi}\, D_\Lambda k^\Xi_a=0\quad \Longrightarrow\quad
h^{(0)}_{\Lambda,b\lambda }\, {\partial (f^{abc} q^{c\lambda})\over \partial q^\Sigma}+ h^{(0)}_{\Sigma,b\lambda }\,{\partial (f^{abc} q^{c\lambda})\over \partial q^\Lambda}=0.
\end{align}
In the above equation, $D$ is now the covariant derivative on $\t\M_H$. The choice $(\Lambda,\Sigma)=(d\rho,e\sigma)$ gives $[h^{(\rho\sigma)},T^a]=0$, where we have defined $h^{(\rho\sigma)}_{db}\equiv h^{(0)}_{d\rho,b\sigma}$, which is therefore proportional to the identity matrix. Moreover, for $(\Lambda,\Sigma)=(d\rho,\alpha)$ one obtains $h^{(0)}_{\alpha,e\rho}=0$. In total, we have at this stage
\begin{align}
\label{h0}
h_{a\lambda,b\rho}=\delta_{ab}h_{\lambda\rho}+\O(q-q_0),\qquad h_{a\lambda,\alpha}=\O(q-q_0), \qquad h_{\alpha\beta}=h_{\alpha\beta}^{(0)}+\O(q-q_0),
\end{align}
where $h_{\lambda\rho}$ is a constant metric.

To characterize the hyper-\Ka forms $K^x$ at $P_0$, we use the fact the quaternionic structure must be preserved by the isometric flow, up to $SU(2)$ transformations. This is formulated by saying that there exist sections $W^x_a$ of the $SU(2)$-bundle over $\t\M_H$, such that
\begin{align}
\label{KW}
\L_aK^x=\epsilon^{xyz}K^yW_a^z,
\end{align}
where $\L_a$ are the Lie derivatives with respect to the Killing vectors $k_a^\Lambda\partial_{q^{\Lambda}}$. The $W^x_a$'s are called compensators and we first determine their value at $P_0$. To do so, we consider the relation \cite{Andrianopoli:1996cm}
\begin{align}
\label{compen}
\L_a\omega^x=\nabla^{SU(2)} W_a^x\equiv dW_a^x+\epsilon^{xyz}\omega^y W^z_a.
\end{align}
Starting from the definition $\L_a\omega^x=d\, i_a\omega^x+i_ad\omega^x$, where $d\omega^x+{1\over 2}\epsilon^{xyz}\omega^y\wedge\omega^z\equiv\Omega^x=\lambda K^x$ is the $SU(2)$ curvature
 proportional to the hyper-\Ka forms, and using $i_a(\omega^y\wedge \omega^z)=i_a\omega^y\, \omega^z-\omega^y\,i_a\omega^z$, we obtain from Eq. (\ref{compen})
 \begin{align}
 \nabla^{SU(2)}W^x_a=\lambda i_aK^x+ \nabla^{SU(2)}i_a\omega^x.
 \end{align}
Combining with Eq. (\ref{momna}), we find
\begin{align}
\nabla^{SU(2)}\left({\lambda\over2}\,\P^x_a+W^x_a-i_a\omega^x\right)=0\qquad \Longrightarrow\qquad {\lambda\over2}\,\P^x_a+W^x_a-i_a\omega^x=0,
\end{align}
by virtue of Theorem 3 in Appendix C. Since both $\P^x_a$ and $k_a^\Lambda\partial_{q^{\Lambda}}$ vanish at $P_0$, we conclude that $W_a^x(q_0)=0$ as well. We are now in a position to use efficiently Eq. (\ref{KW}) we rewrite in components as,
\begin{align}
\forall \Lambda,\Sigma,a:\quad k^\Xi_a\, \partial_{q^\Xi}K^x_{\Lambda\Sigma}+(\partial_{q^\Lambda} k^\Xi_a)K^x_{\Xi\Sigma}+(\partial_{q^\Sigma} k^\Xi_a)K^x_{\Xi\Lambda}=\epsilon^{xyz}K^y_{\Sigma\Lambda}W_a^z.
\end{align}
At $q=q_0$, this relation with $(\Lambda,\Sigma)=(b\lambda,c\rho)$ gives $[K^{x(\lambda\rho)},T^a]=0$, where we have defined $K^{x(\lambda\rho)}_{bc}\equiv K^{x(0)}_{b\lambda,c\rho}$,which is thus proportional to the identity matrix. With  $(\Lambda,\Sigma)=(b\lambda,\alpha)$, one obtains instead $K^{x(0)}_{c\lambda,\alpha}=0$, so that
\begin{align}
\label{K(0)}
K^x_{a\lambda,b\rho}=\delta_{ab}K^x_{\lambda\rho}+\O(q-q_0),\qquad K^x_{a\lambda,\alpha}=\O(q-q_0), \qquad K^x_{\alpha\beta}=K_{\alpha\beta}^{x(0)}+\O(q-q_0),
\end{align}
where $K^x_{\lambda\rho}$ is an antisymmetric constant matrix.

The triplet of complex structures are related to the hyper-\Ka forms by the definition (\ref{KJ}), so that $J^{x\Lambda}{}_\Sigma=-h^{\Lambda \Xi}K^x_{\Xi\Sigma}$. Using Eqs (\ref{h0}) and (\ref{K(0)}), we obtain in the vicinity of $P_0$,
 \begin{align}
\label{J0}
J^{xa\lambda}{}_{b\rho}=\delta^a_bJ^{x\lambda}{}_\rho+\O(q-q_0),\qquad J^{xa\lambda}{}_\alpha=\O(q-q_0), \qquad J^{x\alpha}{}_\beta=J^{x(0)\alpha}{}_\beta+\O(q-q_0),
\end{align}
where $J^{x\lambda}{}_\rho=-h^{\lambda\sigma}K^x_{\sigma\rho}$. Using the fact that the $J^{x\Lambda}{}_\Sigma$'s are satisfying the quaternionic algebra (Eq. (\ref{TRM0}) with $n=h_{12}+1+g(N^2-N)$) everywhere on $\t\M_H$ and thus at $P_0$, we find $J^{x\lambda}{}_\sigma$ are also triplets of complex structures (Eq. (\ref{TRM0}) with $n=g$) in each of the $N^2-1$ sub-tangent planes $\T_{0a}$ at $P_0$ spanned by $\partial/\partial_{q^{a\lambda}}$ (at fixed $a$), endowed with the metric $h_{\lambda\rho}$. Applying Theorem 1 (see Appendix B) in $\T_{0a}$, there exists a new basis $e_{a\A u}$ in $\T_{0a}$ and its dual basis $\theta^{a\A u}$ in $\T^*_{0a}$, where $\A=1,\dots,g$ and $u=1,2,3,4$, such that $J^{x\lambda}{}_\rho$, $h_{\lambda\rho}$ and $K^x_{\lambda\rho}$ take canonical forms. This local basis defines a new set of coordinates $c^{a\A u}$ such that $\left.dc^{a\A u}\right\abs_{P_0}=\sqrt{2}\, \theta^{a\A u}$ and $\left.c^{a\A u}\right\abs_{P_0}=0$, which greatly simplifies the form of the metric, hyper-\Ka forms and Killing vectors on $\t\M_H$. Using Eqs (\ref{h0}), (\ref{K(0)}) and (\ref{k0}), we obtain in the neighborhood of $P_0$,\footnote{$\O(q-q_0)$ denotes terms of order $c^{a\A u}$ or $(q^\alpha-q_0^\alpha)$.}
\begin{align}
\begin{array}{lll}
\dis h_{a\A u,b\B v}={1\over 2}\delta_{ab}\delta_{\A\B}\,  \delta_{uv}+\O(q-q_0), & h_{{a\A u,\alpha}}=\O(q-q_0), & h_{\alpha\beta}=h^{(0)}_{\alpha\beta}+\O(q-q_0),\\
\dis K^x_{a\A u,b\B v}={1\over 2}\delta_{ab}\delta_{\A\B}\, \eta^x_{uv}+\O(q-q_0), & K^x_{{a\A u,\alpha}}=\O(q-q_0), &K^x_{\alpha\beta}=K^{x(0)}_{\alpha\beta}+\O(q-q_0),\esp\\
\label{flat1}k^{b\A u}_a=f^{abc}\, c^{c\A u}+\O((q-q_0)^2),& k^\alpha_a=\O((q-q_0)^2).&
\end{array}
\end{align}

Finally, we know the momentum maps $\P^x_a$ are vanishing at $P_0$. Moreover, since the left hand side of Eq. (\ref{momna}) is first order, the $\P^x_a$'s must be second order. To determine them explicitly in the
neighborhood of $P_0$, we rewrite Eq. (\ref{momna}) as
\begin{align}
{\partial\P^x_a\over \partial c^{b\B v}}=-f^{abc}c^{c\B u}\eta^x_{uv}+\O((q-q_0)^2),\qquad {\partial \P^x_a\over \partial q^\alpha} =\O((q-q_0)^2),
\end{align}
whose solution is
\begin{align}
\label{Pna}
\P^x_a={1\over 2}\, f^{abc}\, c^{b\A u}c^{c\A v}\eta^x_{uv}+\O((q-q_0)^3).
\end{align}

\vskip .2cm
\noindent {\large \em Effective action:} We are now equipped to determine the potential $\V$ of Eq. (\ref{Vcla}) in the vicinity of $P_0$. As follows from Eqs (\ref{gna}), (\ref{knav}), (\ref{flat1}), the first three terms in $\V$, which are positive,  are of order four in $(X-X_0)$ or $(q-q_0)$, while the last one, which is negative, is of order six and can be ignored. Introducing the $SU(2)_R$-doublets and  notations
\begin{align}
\label{douNa}
\cC^{a\A}=\left(\!\!\begin{array}{c} i(c^{a\A 1}+ic^{a\A 2})\\(c^{a\A 3}+ic^{a\A 4})^*\end{array}\!\!\right), \qquad X\equiv X^a\, T^a, \quad \b X\equiv X^a\, T^a,\quad c^{\A u}\equiv c^{a\A u}\,T^a,
\end{align}
the tree level potential can be written as
\begin{align}
\label{V0na}
&\V= e^{\K^{(0)}}\Big(l^2[X,\b X]^a[X,\b X]^a+2[X,c^{\A u}]^a[c^{\A u},\b X]^a+{1\over 4l^2}D^{ax} D^{ax} \Big)+\cdots\nonumber \\
&\where\qquad  D^{ax}\equiv-if^{abc}\, \mfs C^{b\ca\dagger}\sigma^x\mfs C^{c\ca},\qquad [{\cal X},{\cal Y}]^a\equiv if^{abc}\, {\cal X}^b{\cal Y}^c,
\end{align}
with the dots standing for order five terms.

Collecting the  leading contributions to the kinetic terms, the low energy effective action (\ref{Scla}) of the type IIA theory at finite temperature and compactified on $M$ or $M^\dprime$, close to their extremal transition, is at tree level and lowest order in scalar fields,
\begin{align}
\label{Sna}
S_{\rm tree}=\int d^4x \,\sqrt{-g}\,\bigg\{&{\R\over 2}-l^2\, \nabla_{\!\!\mu} X^a\nabla^\mu \b X^a-g^{(0)}_{p\b q}\, \partial_\mu X^p\partial^\mu \b X^q - {1\over 2}\,\nabla_{\!\!\mu} c^{a\A u}\nabla^\mu c^{a\A u} -h^{(0)}_{\alpha\beta}\,\partial_\mu q^\alpha\partial^\mu q^\beta\nonumber\\
&-e^{\K^{(0)}}\Big(l^2[X,\b X]^a[X,\b X]^a+2[X,c^{\A u}]^a[c^{\A u},\b X]^a+{1\over 4l^2}D^{ax} D^{ax}  \Big)+\cdots\bigg\}.
\end{align}
Therefore, up to the ``spectator multiplets", the Lagrangian density has the form of a minimally coupled rigid $\N=2$ supersymmetric $SU(N)$ gauge theory coupled to $g$ hypermultiplets in the adjoint representation, formally coupled to gravity.


\subsection{Tree level masses}
\label{massNA}

The tree level potential (\ref{V0na}) admits flat directions, along which masses for the degrees of freedom involved in the $SU(N)$ gauge theory are generated. Since they depend on the  moduli, the one-loop free energy we will take into account in the following Sections will behave as an effective potential, able to lift the flat directions of the vector and hypermultiplet scalars charged under $SU(N)$. Actually, rather than the masses themselves, this goal requires the sum of the tree level squared masses of the bosonic degrees of freedom we now determine.

Let us start by characterizing the flat directions of $\V$. In the vicinity of  $P_0$, whose coordinates in $\t\M_V\times \t\M_H$ are $(X_0^a=0,X^p_0;c_0^{a\A u}=0, q^\alpha_0)$, the set of vacua splits into a Cartesian product between:
\begin{itemize}

\item The space parameterized by the ``spectator moduli" $X^p$ and $q^\alpha$, which are coordinates along the non-Abelian locus.

\item The space of configurations of the $X^a$'s and $c^{a\A u}$'s, for which the semi-definite positive potential $\V$ vanishes. It is defined by the conditions
\begin{align}
\label{vacNa}
[X,\b X]=0,\qquad \forall \A,u: ~~ [X,c^{\A u}]=0 \qquad \and  \qquad \forall a,x : ~~D^{ax}=0,
\end{align}
and admits Coulomb and Higgs branches to be specified later.
\end{itemize}
To determine the mass terms, we substitute in the action  (\ref{Sna})
\begin{align}
\label{subs}
X^a\to X^a+\delta X^a, \qquad c^{a\A u}\to c^{a\A u}+\delta c^{a\A u},
\end{align}
where $(X^a, c^{a\A u})$ satisfies the constraints (\ref{vacNa}), and focus on the quadratic terms of the potential. For $\delta X^b\delta \b X^c$, we obtain from the first two terms of $\V$ in Eq. (\ref{V0na})
\begin{align}
e^{\K^{(0)}} \Big(2 l^2 [\delta X,\delta \b X]^a[X,\delta\b X]^a+2[\delta X,c^{\A u}]^a[c^{\A u},\delta \b X]^a\Big):=E_{bc}\, \delta X^b\delta \b X^c,
\end{align}
while the contributions in $\delta X^b\delta X^c$ and their complex conjugate are irrelevant to compute the trace  of the squared mass operator.
Taking into account the normalization of the kinetic terms of the $\delta X^a$'s in the action (\ref{Sna}), we find a contribution to the trace
\begin{align}
\label{vna}
\left.\tr M^2\right\abs_{\delta X}={2\over l^2}\, E_{aa}+\cdots=4N e^{\K^{(0)}}\Big( X^a\b X^a+{1\over l^2}\, c^{a\A u} c^{a\A u}\Big)+\cdots.
\end{align}
The factor 2 counts the real and imaginary parts of $\delta X^a$, the factor $N$ arises from the relation $\tr (T^aT^b)=N\delta^{ab}$ and the ellipsis stand for higher order terms in the scalar fields.

Terms in $\delta c^{a\A u}\delta c^{b\B v}$ arise from the second term in (\ref{V0na}),
\begin{align}
e^{\K^{(0)}} 2 [X,\delta c^{\A u}]^a[\delta c^{\A u},\b X]^a:= {1\over 2}\, L_{a\A u,b\B v}\, \delta c^{a\A u}\delta c^{b\B v},
\end{align}
and the D-terms. To evaluate the latter, it is convenient to rewrite  the third term of Eq. (\ref{Vcla}) around $P_0$ in an alternative form, using Eq. (\ref{thft4}),
\begin{align}
\label{DDna}
e^{\K^{(0)}} {1\over 4 l^2}\, D^{ax}D^{ax}=e^{\K^{(0)}} {1\over 4 l^2}\Big(2[c^{\A v},c^{\A u}]^a[c^{\B u},c^{\B v}]^a-\epsilon_{uvu'v'}[c^{\A u},c^{\A v}]^a[c^{\B u'},c^{\B v'}]^a\Big).
\end{align}
Substituting (\ref{subs}), the only terms that may contribute  to the trace of squared masses are
\begin{align}
\label{cna}
e^{\K^{(0)}}{1\over l^2} \Big([\delta c^{\A v},c^{\A u}]^a [\delta c^{\B u},c^{\B v}]^a+[\delta c^{\A v},c^{\A u}]^a [c^{\B u},\delta c^{\B v}]^a\Big):= {1\over 2}\, P_{a\A u,b\B v}\, \delta c^{a\A u}\delta c^{b\B v}.
\end{align}
The kinetic terms of the $\delta c^{a\A u}$'s being canonical, we obtain
\begin{align}
\label{hna}
\left.\tr M^2\right\abs_{\delta c}=(L+P)_{a\A u,a\A u}+\cdots=4N e^{\K^{(0)}}\Big( 4gX^a\b X^a+{3\over 2l^2}\, c^{a\A u} c^{a\A u}\Big)+\cdots.
\end{align}
All terms in $\delta X\delta c^{a\A u}$ or their complex conjugate do not contribute the trace.  This completes the analysis arising from the scalar fields.

To proceed, we need the contribution associated to the vector bosons. The Yang-Mills Lagrangian density implicit in Eq. (\ref{Sna}) can be written as
\begin{align}
-{1\over 4}\, \tau^{(0)}_{AB}\, F^A_{\mu\nu}F^{B\mu\nu},
\end{align}
where the index  $A$ takes values 0 or $I$ to account for the graviphoton, $\tau^{(0)}_{AB}$ are the gauge kinetic functions evaluated at $P_0$ and $F^A=dA^A$. When (\ref{subs}) is applied and we keep the lowest order terms in scalar fields, a mass matrix $Q_{ab}$ is  generated by the covariant derivatives and we obtain in flat Minkowski space,
\begin{align}
{1\over 2}\, \tau^{(0)}_{AB}\, A^{A\mu}    (\eta_{\mu\nu}\Box-\partial_\mu\partial_\nu)     A^{B\nu}-{1\over 2}\, A^{a\mu}\, \eta_{\mu\nu}Q_{ab}\, A^{b\nu}.
\end{align}
Therefore, the contribution of the vector bosons to the trace of squared masses is
\begin{align}
\label{MA}
\left.\tr M^2\right\abs_{A^\mu}=3\, \tau^{(0)ba}Q_{ab}+\cdots=6\, \tau^{(0)ba}f^{cda}f^{ceb}\Big( l^2 X^d\b X^e+{1\over 2}\, c^{d\A u} c^{e\A u}\Big)+\cdots,
\end{align}
where the factor 3 counts the number of degrees of freedom in massive spin-one fields. It is certainly possible to evaluate the inverse metric $\tau^{(0)ab}$ from its form dictated by $\N=2$ supergravity \cite{Andrianopoli:1996cm}. However, this can be avoided by noticing that $\left.\tr M^2\right\abs_{A^\mu}$ is known when $g=1$. In this conformal case, there is no Higgs branch and, in the Coulomb phase, all massive degrees of freedom sit in long vector multiplets (see Table \ref{table na}). Each of them contains 3 degrees of freedom associated to the vector bosons and 5 from the scalars, which are all degenerate. Therefore, we have
\begin{align}
\left.\tr M^2\right\abs_{A^\mu}^{g=1}={3\over 5}\, \Big( \! \left.\tr M^2\right\abs_{\delta X}^{g=1}+   \left.\tr M^2\right\abs_{\delta c}^{g=1}\!\Big)+\cdots=12N e^{\K^{(0)}}\Big( X^a\b X^a+{1\over 2l^2}\, c^{a1 u} c^{a1 u}\Big)+\cdots.
\end{align}
Comparing with Eq. (\ref{MA}), we find $\tau^{(0)ba}f^{cda}f^{ceb}l^2\equiv 2N e^{\K^{(0)}}\delta^{de}$ and therefore for arbitrary $g\ge 1$,
\begin{align}
\label{MA2}
\left.\tr M^2\right\abs_{A^\mu}=12N e^{\K^{(0)}}\Big( X^a\b X^a+{1\over 2l^2}\, c^{a\A u} c^{a\A u}\Big)+\cdots.
\end{align}

Collecting the contributions of the scalar masses Eqs (\ref{vna}), (\ref{hna}) and spin-one fields Eq. (\ref{MA2}), we arrive at the final result for the trace of the squared mass operator in the bosonic sector of the $SU(N)$ gauge theory coupled to $g$ hypermultiplets in the adjoint representation, valid in the vicinity of $P_0$,
\begin{align}
\label{mna}
\left.\tr M^2\right\abs_{\rm gauge}=16N e^{\K^{(0)}}\Big( (g+1)X^a\b X^a+{1\over l^2}\, c^{a\A u} c^{a\A u}\Big)+\cdots.
\end{align}
We recall that the above expression applies to any scalar configuration satisfying (\ref{vacNa}), in which case all tadpoles are vanishing.


\subsection{Lifting the Coulomb branch at one-loop}
\label{nonACoul}

The space of classical flat directions around $P_0$ splits into several branches. The Coulomb phase corresponds to scalar VEV's such that the matrices $X^aT^a$ and $c^{a\A u}T^a$ sit in the Cartan sub-algebra. In this case, the two first conditions in Eq. (\ref{vacNa}) are satisfied, while the D-term condition follows from Eq. (\ref{DDna}). Denoting as $T^i$ ($i=1,\dots, N-1$) the Cartan generators of $SU(N)$ and $T^\ha$ ($\ha=N,\dots,N^2-N$) the remaining ones, we have
\begin{align}
\label{coulNA}
\mbox{Coulomb branch:}\,\Big\{\big(X^i \mbox{ arbitrary}, X^\ha=0,c^{i\A u} \mbox{ arbitrary}, c^{\ha\A u}=0\big)\Big\}\times\Big\{\big(X^p, q^\alpha\big) \mbox{ arbitrary}\Big\},
\end{align}
which corresponds to a compactification on the CY space $M$. As noticed in Footnote \ref{SU}, $M$ is  a smooth manifold, except when some $X^i=X^j$ and $c^{i\A u}=c^{j\A u}$  for $i\neq j$, so that $SU(N)$ is broken to a non-Abelian subgroup of rank $N-1$.

In the coulomb phase, when the $X^i$'s are generic but $c^{i\A u}=0$ (for all $\A$ and $u$), no mass mixing terms of the form $\delta X^a\delta c^{b\B v}$ or $\delta \b X^a\delta c^{b\B v}$ are generated by the shift (\ref{subs}) in $\V$. Therefore, the $N^2-N$ gauge bosons that are acquiring masses eat half of the degrees of freedom of the $\delta X^\ha$'s  and  $\delta \b X^\ha$'s, and combine into $N^2-N$ short massive vector multiplets. Thus, the ratio $\left.\tr M^2\right\abs_{A^\mu}/\left.\tr M^2\right\abs_{\delta X}$ must equal $3/(2-1)=3$, which is satisfied by Eqs (\ref{MA}) and (\ref{vna}). The complete superfield content in this case is reported in Table \ref{table na}.
\begin{table}[h]
	\centering
\scalebox{0.84}{
	   \begin{tabular}{|c|c|c|c|c|c|c|c|}
		\cline{2-8}
		     \multicolumn{1}{c|}{} & \multicolumn{2}{c}{\text{Scalars acquiring VEV's}} & \multicolumn{5}{|c|}{\text{Superfields}} \\ \cline{2-8}
              \multicolumn{1}{c|}{} & \raisebox{-0.1cm}{In vector} & \raisebox{-0.1cm}{In hyper-} & \multicolumn{3}{c|}{Vector multiplets} & \multicolumn{2}{c|}{Hypermultiplets}\\ \cline{4-8}
                \multicolumn{1}{c|}{} &\raisebox{0.2cm}{multiplets}  & \raisebox{0.2cm}{multiplets} & $\substack{\text{Massless}\\ \text{(moduli)}}$ & $\substack{\text{Massive}\\ \text{short}}$ & $\substack{\text{Massive}\\ \text{long}}$ & $\substack{\text{Massless}\\ \text{(moduli)}}$ & \small{Massive} \\ \cline{1-8}  & & & & & & & \\
            \raisebox{-0.5cm}{Coulomb} & $X^i$ & none & $N-1$ & $N^2-N$ & $0$ & $g(N-1)$ & $g(N^2-N)$\\  \cline{2-8}
              $\substack{\\ \text{\small phase}}$& & & & & & & \\
           & $\begin{array}{c} X^i\\ \text{or none} \end{array}$ & $c^{i\A u}$ & $N-1$ & 0 & $N^2-N$ & $g(N-1)$ & $(g-1)(N^2-N)$ \\ & & & & & & & \\  \hline
          $\begin{array}c \mbox{Higgs} \\\mbox{phase}\end{array}$ & none & \multicolumn{1}{m{2.2cm}|}{$\mfs C^{a\ca}$ mod. gauge orbits such that $D^{ax}=0$} & 0 & 0 & $N^2-1$ & $(g-1)(N^2-1)$ & 0
        \\ \hline
	   \end{tabular}
	   }
	\caption{\small Superfield contents in the Coulomb and Higgs branches (when $g\ge 2$) associated to the $\N=2$ $SU(N)$ gauge theory coupled to $g$  hypermultiplets in the adjoint representation, which is  encountered in the neighborhood of a non-Abelian locus in $\t \M_V\times \t\M_H$. The  scalars $X^p$ and $q^\alpha$ of the massless spectator vector multiplets and hypermultiplets  are not represented. At special loci in the Coulomb branch, where some $X^i=X^j$ and $c^{i\A u}=c^{j\A u}$  for $i\neq j$, some generically massive multiplets are actually massless, and the $SU(N)$ gauge symmetry is broken to a non-Abelian subgroup of rank $N-1$, rather than $U(1)^{N-1}$.}
	\label{table na}
\end{table}

Similarly, when $X^i=0$ (for all $i$) but the $c^{i\A u}$'s are generic, there are still no mixing terms between the vector and matter scalars. However, the $N^2-N$ vector boson eat as many would-be-Goldstone bosons now among the $\delta c^{\ha \A u}$, and  the massive spectrum contains $N^2-N$ long vector multiplets. Thus, the ratio $\left.\tr M^2\right\abs_{A^\mu}/\left.\tr M^2\right\abs_{\delta X}$ must equal $3/2$, again satisfied by Eqs (\ref{MA}) and (\ref{vna}).

On the contrary, at a generic point in the branch (\ref{coulNA}), the mass mixing terms imply the mass eigenstates are combinations of vector and hypermultiplet scalars.  Therefore, $\left.\tr M^2\right\abs_{\delta X}$ and $\left.\tr M^2\right\abs_{\delta c}$ cannot be interpreted as sums of squared masses in separated vector and hypermultiplet scalar sectors. Only the sum  $\left.\tr M^2\right\abs_{\delta X}+\left.\tr M^2\right\abs_{\delta c}$ makes sense, as the contribution of the full spin-zero sector. Moreover, the would-be-Goldstone bosons are combinations of the non-Cartan vector and hypermultiplet scalars.

Restricted to the weak string coupling regime, we are now ready to write the one-loop thermal effective action. In the Coulomb branch, it amounts to adding the tree level action (\ref{Sna}) in a vacuum (\ref{coulNA}) to the one-loop Coleman-Weinberg effective potential $\F$,
\begin{align}
\label{Sna1}
S_{\mbox{\scriptsize 1-loop}}\!=\!\!\int \!d^4x \sqrt{-g}\bigg\{{\R\over 2}\!-\!l^2 \partial_\mu X^i\partial^\mu \b X^j \!-\!g^{(0)}_{p\b q} \partial_\mu X^p\partial^\mu \b X^q\! -\! {1\over 2}\partial_\mu c^{a\A u}\partial^\mu c^{a\A u} \!-\!h^{(0)}_{\alpha\beta}\partial_\mu q^\alpha\partial^\mu q^\beta\!-\!\F\bigg\}\!.
\end{align}
As seen in Eq. (\ref{freegene}), $\F$ is actually the free energy density, which in the present case is
\begin{align}
\label{FNAc}
\F=-T^4 \left\{ \Big(4+ 4h_{11}+4(h_{12}+1)\Big)G(0) +\sum_{\h s} G\Big({M_{\h s}\over T}\Big)+\O\big(e^{-{M_{\rm min}\over T}}\big)\right\}\!,
\end{align}
where the index $\h s$ labels all pairs of degenerate boson/fermion states in the massive vector multiplets and hypermultiplets involved in the $SU(N)$ gauge theory and collected in Table  \ref{table na}. In Eq. (\ref{FNAc}), we take the temperature to be below the lower bound $M_{\rm min}>0$ at $P_0$ of the remaining masses of the full string spectrum. $\F$ is minimal when and only when all classical masses in the $SU(N)$ gauge sector vanish, $\forall \h s:\; M_{\h s}=0$. Using the general formula (\ref{mna}) in the Coulomb branch,
\begin{align}
\label{mnac}
\sum_{\h s} M^2_{\h s}\equiv \left.\tr M^2\right\abs_{\rm gauge}=16N e^{\K^{(0)}}\Big( (g+1)X^i\b X^i+{1\over l^2}\, c^{i\A u} c^{i\A u}\Big)+\cdots,
\end{align}
this implies $X^i=0$, $c^{i\A u}=0$. Therefore, all moduli involved in the Coulomb phase of the $SU(N)$ gauge theory are lifted. Their kinetic terms being diagonal, the masses they acquire at one-loop are
\begin{align}
&M^2_{i\mbox{\scriptsize 1-loop}}=\left.{1\over l^2}\, {\partial^2\F\over \partial X^i\partial X^i}\right\abs_{X^j=c^{j\A u}=0}={T^2\over 16}\,  {1\over l^2} \left.\sum_s{\partial^2M_s^2\over \partial X^i\partial X^i}\right\abs_{X^j=c^{j\A u}=0}=T^2\, (g+1)\, {N\over l^2}e^{\K^{(0)}},\nonumber \\
&M^2_{i\A u\mbox{\scriptsize 1-loop}}=\left.{\partial^2\F\over \partial c^{i\A u}\partial c^{i\A u}}\right\abs_{X^j=c^{j\A u}=0}={T^2\over 16} \left.\sum_s{\partial^2M_s^2\over \partial c^{i\A u}\partial c^{i\A u}}\right\abs_{X^j=c^{j\A u}=0}=T^2\, 2\, {N\over l^2}e^{\K^{(0)}},
\end{align}
while the classically massless $U(1)^{N-1}$ spin-1 fields do not acquire masses. Moreover, due to the arbitrariness in the choice of Cartan subalgebra at the origin of the Coulomb branch, we conclude that all vector multiplet and hypermultiplet scalars $X^a$ and $c^{a\A u}$, even though classically massless, have one-loop masses at their point of stabilization,
\begin{align}
\label{Xc}
M^2_{X\mbox{\scriptsize 1-loop}}=T^2\, (g+1)\, {N\over l^2}e^{\K^{(0)}},\qquad M^2_{c\mbox{\scriptsize 1-loop}}=T^2\, 2\, {N\over l^2}e^{\K^{(0)}},
\end{align}
where the full $SU(N)\times U(1)^{h_{11}-(N-1)+1}$ gauge symmetry is restored.

In a homogeneous and isotropic universe, the free energy density (\ref{FNAc}) leads along the $SU(N)$ non-Abelian locus to the state equation for radiation, $\rho=3P$, when $T$ is low enough. Consequences of this fact are similar to those encountered in the case of the conifold locus, below Eq. (\ref{rhoP}). A cosmological evolution for the scale factor and temperature exists, with static scalars fields,
\begin{align}
\label{solpartNA}
a(t)\propto {1\over T(t)}\propto \sqrt{t}\; , \quad X^a\equiv c^{a\A u}\equiv 0\; , \quad X^p, q^\alpha\, \mbox{constant},
\end{align}
where $t$ is the cosmological time. Since $T$ decreases, the consistency of the approximation $T\ll M_{\rm min}$ used to neglect exponentially suppressed contributions in the free energy (\ref{FNAc}) is guaranteed. As in the $\N=4$ case ($g=1$ here) analyzed in Ref. \cite{Estes:2011iw}, other time-evolutions compatible with homogeneity and isotropy exist, where the moduli $X^i$ and $c^{i\A u}$ oscillate with damping  in the Coulomb branch, thus converging to their minimum. The fact that their masses are of order the temperature scale and decreases as the universe expand implies the cosmological moduli problem is avoided.


\subsection{Lifting the Higgs branch at one-loop}
\label{nonAHiggs}

At the origin of the Coulomb branch, the conditions $D^{ax}=0$ become non-trivial constraints on the hypermultiplet scalars, which define the Higgs phase,
\begin{align}
\label{higgsNA}
\mbox{Higgs branch : } \quad \Big\{\big(X^a =0, \cC^{a\A}\mbox{ such that } D^{bx}=0\big)\Big\}\times\Big\{\big(X^p, q^\alpha\big) \mbox{ arbitrary}\Big\}.
\end{align}
The above conditions fix $3(N^2-1)$ components among the $4g(N^2-1)$ scalars $c^{a\A u}$. Along the  flat directions, the $SU(N)$ local symmetry is completely Higgsed spontaneously. The remaining global $SU(N)$ orbits can be used to gauge away $N^2-1$ would-be-Goldstone bosons, so that $4(g-1)(N^2-1)$ flat directions of inequivalent vacua remain. By supersymmetry, the latter can be parameterized by the scalars of $(g-1)(N^2-1)$ neutral hypermultiplets. Thus, the above Higgs branch exists only for $g\ge 2$, in which case it is realized geometrically by compactifying on the CY $M^\dprime$ with Hodge numbers given in Eq. (\ref{h''}). Actually, $N^2-1$ of the initial $g(N^2-1)$ hypermultiplets combine with the Higgsed vector multiplets into $N^2-1$ massive long vector multiplets, as summarized in Table \ref{table na}. Therefore  $\left.\tr M^2\right\abs_{A^\mu}/\left.\tr M^2\right\abs_{\delta X}=\left.\tr M^2\right\abs_{\delta c}/\left.\tr M^2\right\abs_{\delta X}=3/2$ must be satisfied, which is consistent with Eqs (\ref{MA}), (\ref{vna}) and (\ref{hna}) when $X^a=0$.

In the case of the Coulomb phase, we introduced an arbitrary choice of Cartan generators $T^i$ among the $T^a$'s.  In a similar way, we now define an arbitrary set of coordinates $\xi^m$ \big($m=1,\dots,4(g-1)(N^2-1)$\big) along the Higgs phase flat directions. They satisfy $f^{abc}\, \mfs C^{b\ca\dagger}(\xi)\sigma^x\mfs C^{c\ca}(\xi)=0$ and the Jacobian matrix $\Big({\partial c^{a\ca u}\over \partial \xi^m}\Big)$ is of rank $4(g-1)(N^2-1)$. The origin of the Higgs branch is denoted $\xi_0^m$. In these notations, the one-loop effective action of the type IIA string theory compactified on $M^\dprime$ at finite temperature is, in the neighborhood of $P_0$,
\begin{align}
\label{SHNA}
S_{\mbox{\scriptsize 1-loop}}=\int d^4x \, \sqrt{-g}\,\bigg\{{\R\over 2}  -g^{(0)}_{p\b q}\,\partial X^p\partial \b{X}^q -h_{mn}^{(0)} \, \partial \xi^m \partial \xi^n-h^{(0)}_{\alpha\beta}\, \partial q^{\alpha} \partial q^{\beta} -\F\bigg\},
\end{align}
where the induced metric of the $\xi^m$'s is
\begin{align}
 h_{mn}^{(0)} ={1\over 2}\!\left.{\partial c^{a\A u} \over \partial \xi^m}\right\abs_{\xi_0}\left.{\partial c^{a\A u} \over \partial \xi^n}\right\abs_{\xi_0}.
 \end{align}
In the present case, the  free energy density $\cf$ takes the form
\begin{align}
\label{FHNA}
\F=-T^4 \left\{ \Big(4+ 4h^\dprime_{11}+4(h^\dprime_{12}+1)\Big)G(0) +8\sum_a G\Big({M_a\over T}\Big)+\O\big(e^{-{M_{\rm min}\over T}}\big)\right\}\!,
\end{align}
where the factor 8 counts the number of boson/fermion pairs of states in the long  vector multiplets of masses $M_a$ ($a=1,\dots, N^2-1$). The contributions of all the other massive modes of the spectrum are exponentially suppressed, when $T<M_{\rm min}$.

The minimum of $\F$ is reached when the classical masses $M_a$ vanish. Applying  Eq. (\ref{mna}) in the Higgs branch,
\begin{align}
\label{mHna}
8\sum_a M_a^2\equiv \left.\tr M^2\right\abs_{\rm gauge}=16{N\over l^2} e^{\K^{(0)}}\, c^{a\A u} c^{a\A u}+\cdots,
\end{align}
we see that this selects the origin of the Higgs branch, $c^{a\A u}=0$. Therefore, all classical flat directions $\xi^m$ are lifted and admit a unique minimum at $\xi^m_0$ at the one-loop level. The squared mass matrix of the $\xi^m$'s is
\begin{align}
\label{masss}
{\Lambda^{\dprime m}}_n={1\over 2}h^{(0)ml}\left.{\partial^2\F\over \partial\xi^l\partial\xi^n}\right\abs_{\xi_0}={T^2\over 16}\left. {1\over 2}h^{(0)ml} \, 8\sum_a{\partial^2 M^2_a\over \partial\xi^l\partial\xi^n}\right\abs_{\xi_0}=T^2\, 2\, {N\over l^2}e^{\K^{(0)}}\delta^m_n,
\end{align}
where we have used the fact that $c^{a\A u}\abs_{\xi_0}=0$ to reach the last equality. Thus, the $\xi^m$'s are mass eigenstates and degenerate. Since the parametrization of the Higgs branch was chosen arbitrarily, we obtain that all scalars $c^{a\A u}$ acquire a common mass given by Eq. (\ref{masss}). Consistently, this is the result we already found by approaching the $SU(N)$ non-Abelian locus from the Coulomb branch, Eq. (\ref{Xc}).

From a cosmological point of view, the expanding universe with static moduli and filled with radiation of  Eq. (\ref{solpartNA}) appears as a particular case of a second class of homogeneous and isotropic time-evolutions. In this class, even if the $\xi^m$'s oscillate with damping, the cosmological moduli problem is avoided \cite{Estes:2011iw}.


\section{Stabilization at intersections of extremal transition loci}
\label{inters}

In the previous Sections, we have shown that the thermal effective potential admits local minima along submanifolds in moduli space, where the internal CY develops singularities. Therefore, the intersection points of various such loci are expected to define dynamically preferred configurations of the internal space. In the following, we illustrate this fact on an example in type IIA (IIB), where the \Ka (complex structure) moduli space is completely lifted, together with some of the complex structure (K\"ahler) moduli. This implies in particular that the  axio-dilaton field of the heterotic dual description is stabilized. We then discuss how this phenomenon is expected to apply in generic CY compactifications to all vector multiplet and most of the hypermultiplet scalars.


\subsection{Example}
\label{ex}

Let us consider an example, with small Hodge number $h_{11}$ \cite{TwoParam,ho}. The type IIA model we analyze at finite temperature is compactified on a CY manifold $M$ obtained by resolving the singularities of a degree 12 hypersurface in $\CP^4_{(1,1,2,2,6)}$. Denoting the projective coordinates as $x_1,\dots, x_5$, the ambient space presents initially a singularity of type $A_1$ at $x_1=x_2=0$. Along this locus, the polynomial of $M$ defines a curve $\C$ of genus 2. Blowing up the ambient space singularity, $M$ becomes a smooth CY manifold, whose  \Ka moduli space $\M_V$ admits a non-Abelian locus with $N=2$ and $g=2$. Counting the allowed monomials of the defining polynomial of $M$, one finds there are $h_{12}=128$ complex structure moduli.

Equivalently, the model can be analyzed in type IIB compactified on the mirror CY threefold $W$. The latter is  defined by the vanishing locus of degree 12 polynomials   in $\CP^4_{(1,1,2,2,6)}$, modded out by some particular $\Z_6^2\times \Z_2$ group. The most general hypersurface consistent with this action is \cite{TwoParam,Katz,KlemmMyr,ho}
\begin{align}
\label{X12}
\cP=x_1^{12}+x_2^{12}+x_3^6+x_4^6 +x_5^2- 12\psi\, x_1 x_2 x_3 x_4 x_5-2\phi \, x_1^6 x_2^6\, ,
\end{align}
which admits 2 complex structure deformations, $\psi$ and $\phi$. Therefore, the manifold $M$ admits $h_{11}=2$ \Ka moduli.  Defining
\begin{align}
\label{z1z2}
z_1=-{1\over 864}{\phi\over \psi^6}\,,\qquad z_2={1\over \phi^2}\, ,
\end{align}
the locus $\cP =0$ develops singularities when $\Delta_c\Delta_{nA}$ vanishes, where
\begin{align}
\label{Delta}
\Delta_c\equiv (1-z_1)^2-z_1^2 z_2\, ,\qquad \Delta_{nA}\equiv 1-z_2\, .
\end{align}
When $\Delta_c=0$, nodes are occurring, which are identified under $\Z_6^2\times \Z_2$. Therefore, $\M_V$ admits a conifold locus characterized by $R=S=1$.\footnote{We use the fact that the mirror of the conifold $W$ is a conifold $M$.} When $\Delta_{nA}=0$, other isolated singularities occur, yielding again a single singular point on the quotient. Since we know $\M_V$ admits a non-Abelian locus, this second point-like singularity must be associated to the type IIB realization of the $SU(2)$ gauge theory. This is confirmed once the orbifold singularities implied by the discrete modding are blown up to obtain a smooth manifold $W$ \cite{TwoParam,ho}.

Since $R=S$, there is no extremal transition associated to the conifold locus. On the contrary, when $M$ develops the genus-$g$ curve of $A_1$ singularities, the fact that $g>1$ implies that $M$ can be deformed into a distinct smooth CY $M^\dprime$. The ambient spaces, degrees of polynomials and Hodge numbers of the families of CY manifolds on either side of the associated non-Abelian extremal transition are \cite{Katz,KlemmMyr}
\begin{align}
\label{fam}
\CP^4_{(1,1,2,2,6)}[12](2,128)\; \longleftrightarrow\; \CP^5_{(1,1,1,1,1,3)}[2,6](1,129).
\end{align}
Beside the $U(1)_{\rm grav}$ gauge factor associated to the graviphoton, the type IIA compactifications on $M$ and $M^\dprime$ realize geometrically the phases of the $U(1)_{\rm con}$ Abelian theory coupled to a charged hypermultiplet and $SU(2)$ super-Yang-Mills theory coupled to two hypermultiplets in the adjoint representation,
\begin{eqnarray}
\mbox{type IIA on $M\phantom{{}^\dprime}$ :} &&\Big\{ \mbox{Coulomb phase of $U(1)_{\rm con}$}\Big\} \times \Big\{ \mbox{Coulomb phase of $SU(2)$}\Big\}\nonumber\\
\mbox{type IIA on $M^\dprime$ :}& &\Big\{ \mbox{Coulomb phase of $U(1)_{\rm con}$}\Big\} \times \Big\{ \mbox{Higgs phase of $SU(2)$}\Big\}.
\end{eqnarray}

The conifold and non-Abelian loci  intersect at two points on the compactified moduli space $\M_V$\cite{TwoParam},
\begin{align}
\label{inter}
(z_1,z_2)=(1/2,1)\quad \mbox{or} \quad  (\infty,1).
\end{align}
In either of these configurations, the node and isolated singularity on the hypersurface $\cP=0$ are separated from each other and generate independent massless states. From the type IIA point of view on the singular space $M$, the single massless  black hole hypermultiplet has charges $Q_1=1$ with respect to $U(1)_{\rm con}$ and $Q_2=0$ with respect to the $U(1)$ Cartan  subgroup of $SU(2)$. Similarly, the vector multiplet and $g=2$ hypermultiplets in the adjoint representation of $SU(2)$ are neutral with respect to $U(1)_{\rm con}$. Therefore, we can combine the results of the previous Sections and consider the extended moduli space $\t\M_V\times \t\M_H$, which takes into account the scalar fields of the light non-perturbative states arising when $(z_1,z_2)$ is close to one of the critical values in Eq. (\ref{inter}). Choosing a point $P_0$ located at the intersection of the conifold and non-Abelian loci in $\t\M_V\times \t\M_H$, the tree level low energy effective action of the type IIA string theory compactified on $M$ or $M^\dprime$ takes near $P_0$ the form,
\begin{align}
\label{Sna2}
S_{\rm tree}=\int d^4x \,\sqrt{-g}\,\bigg\{&{\R\over 2}-l_1^2\, \partial_\mu X^1\partial^\mu \b X^1-l^2\, \nabla_{\!\!\mu} X^a\nabla^\mu \b X^a \nonumber \\
&- {1\over 2}\,\nabla_{\!\!\mu} c^{1 u}\nabla^\mu c^{1 u}- {1\over 2}\,\nabla_{\!\!\mu} c^{a\A u}\nabla^\mu c^{a\A u} -h^{(0)}_{\alpha\beta}\,\partial_\mu q^\alpha\partial^\mu q^\beta\nonumber\\
&-e^{\K^{(0)}}\Big(2 \,\abs X^1\abs^2 \, c^{1u}c^{1u}+ {1\over 4l_1^2}\,(c^{1u}c^{1u})^2 \Big)\nonumber\\
&-e^{\K^{(0)}}\Big(l^2[X,\b X]^a[X,\b X]^a+2[X,c^{\A u}]^a[c^{\A u},\b X]^a+{1\over 4l^2}D^{ax} D^{ax}  \Big)+\cdots\bigg\}.
\end{align}
Our conventions are as follows: $X^1$ is the  scalar partner of the $U(1)_{\rm con}$ gauge boson and $X^a$ ($a=2,3,4$) is in the adjoint of $SU(2)$. Similarly, $c^{1u}$ are the components of the black hole hypermultiplet, while $c^{a\A u}$ are those of the two   hypermultiplets in the adjoint of $SU(2)$. The  scalars of the 127 (see below for the counting) hypermultiplets that are neutral with respect to $U(1)_{\rm con}\times SU(2)$ are denoted $q^\alpha$ and the metric in their subspace in $\t\M_H$ at $P_0$ is  $h^{(0)}_{\alpha\beta}$. Similarly, $l_1^2$, $l^2$ are  the non-vanishing entries of the \Ka metric on $\t\M_V$ at $P_0$, whose coordinates are  $(X_0^1=0,X_0^a=0;c^{1u}_0=0,c^{a\A u}_0=0,q^\alpha_0)$, so that the \Ka potential defined in Eq. (\ref{Kp}) reduces to $\K^{(0)}=-\ln[i(F_0-\b F_0)]$.

Taking into account one-loop corrections, the scalars $X^1, X^a, c^{1u}$ and $c^{a\A u}$ are stabilized at zero and acquire masses of order the temperature scale, while the $q^\alpha$'s remain flat directions of the thermal effective potential. Moreover, the full $U(1)_{\rm grav}\times U(1)_{\rm con}\times SU(2)$ gauge theory is restored. From a geometrical point of view, starting from a type IIA compactification on $M$, the quantum/thermal effects on the perturbative moduli imply:

$\bullet\phantom{.}$The $h_{11}=2$  \Ka moduli are stabilized in one of the two minima given in Eq. (\ref{inter}).

$\bullet\phantom{.}$The scalars of $g(N-1)=2$ hypermultiplets  are stabilized at the origin of the Coulomb branch of $SU(2)$ in $\M_H$.

$\bullet\phantom{.}$The scalars of the $h_{12}+1-g(N-1)=127$ left-over hypermultiplets remain flat directions in $\M_H$.

\noindent Similarly, starting from a type IIA compactification on $M^\dprime$, the thermal free energy density implies:

$\bullet\phantom{.}$The $h^\dprime_{11}=1$ complexified   \Ka modulus parameterizing $\M_V^\dprime$ is stabilized.

$\bullet\phantom{.}$The scalars of $(g-1)(N^2-1)=3$ hypermultiplets  are stabilized at the origin of the Higgs branch of $SU(2)$ in $\M^\dprime_H$.

$\bullet\phantom{.}$The scalars of the $h^\dprime_{12}+1-(g-1)(N^2-1)=127$ left-over hypermultiplets remain flat directions in $\M^\dprime_H$.


\vskip .2cm
\noindent {\large \em The heterotic dual:}
At zero temperature, the type IIA string model compactified on $M$ admits a heterotic dual description \cite{DualConst}. This follows from the fact that the CY manifolds in the family $\CP^4_{(1,1,2,2,6)}[12]$ are $K3$-fibrations \cite{KlemmMayrK3}. The heterotic model is compactified on $K3\times T^2$, where the 2-torus moduli $T_h$ and $U_h$ are identified, $T_h\equiv U_h$ (their difference being projected out), and the full non-Abelian gauge group is Higgsed. Consistently, the massless spectrum contains 2 vector multiplets associated to the $T_h$ and $S_h$ moduli, where $S_h$ is the heterotic axio-dilaton, together with 129 neutral hypermultiplets.

$T_h$ and $S_h$ are special coordinates that can be identified with those obtained by inverting the mirror map: $z_1(t_1,t_2)$, $z_2(t_1,t_2)$. To render the identification precise \cite{DualConst}, one observes that in the large complex structure limit  of $W$, $t_2\to +\infty$, one finds
\begin{align}
\label{mirmap}
z_1={1728\over j(t_1)}+\cdots,\qquad z_2=e^{-t_2}+\cdots,
\end{align}
where $j$ is the $SL(2,\Z)$-invariant modular form. Therefore, $z_2\to 0$ and the two roots of the discriminant locus $\Delta_c$ in Eq. (\ref{Delta}) merge into $z_1=1$. These facts match exactly the behavior of the perturbative heterotic model, under the identification
\begin{align}
T_h\equiv t_1,\qquad S_h\equiv t_2.
\end{align}
Actually, the latter develops an $SU(2)$ enhanced gauge symmetry\footnote{To not be confused with the $SU(2)$ gauge group occurring at the type II non-Abelian locus.}, when $T_h=i$ modulo the classical T-duality group $SL(2,\Z)$, in perfect agreement with Eq. (\ref{mirmap}) for $z_1=1$. Moreover, when $t_2$ is finite, the conifold locus splits into two branches, as predicted by the exact  pure $SU(2)$ $\N=2$  super-Yang-Mills theory \cite{Seiberg:1994rs}. Being asymptotically free, the latter reduces in the IR to a $U(1)$ gauge theory  coupled to a single (dyonic) hypermultiplet, realized as $U(1)_{\rm con}$ in the type II setup \cite{SWinString,X8}.

In their exact versions, the type II and heterotic models are supposed to be equivalent. Therefore, switching on finite temperature on both theories must lead to a new dual pair of non-supersymmetric models. This expectation is confirmed by the fact that at the levels of the worldsheet conformal field theories, finite temperature is introduced by implementing spontaneous breakings of supersymmetry ``à la Scherk-Schwarz", along the Euclidean time circles. Using an adiabatic argument \cite{Vafa:1995gm}, under a free action, the two theories remain dual. Therefore, the stabilization of the complex structure moduli $z_1,z_2$ of $W$ at one of the two points in Eq. (\ref{inter}) translates immediately into a stabilization of the torus modulus $T_h$ and axio-dilaton $S_h$ in the dual heterotic model at finite temperature. The latter are given by the inverse mirror map, $T_h(z_1,z_2)$, $S_h(z_1,z_2)$, where $z_1=1/2$ or $\infty$ and $z_2=1$. As seen in Eq. (\ref{mirmap}), the obtained value of $S_h$  corresponds to a strong coupling regime of the heterotic theory.

Actually, the two local minima of $(T_h, S_h)$ are uniquely determined, modulo the orbit of the exact heterotic duality group. Since the complex structure moduli space of $W$ is exactly known at tree level in type IIB, the exact heterotic duality group is nothing but the monodromy group derived around the singular loci of the type IIB complex structure moduli space $\M_V$. As shown in Ref. \cite{X8}, the latter contains the perturbative heterotic duality group (including the quantized axionic shift), the monodromies of the exact pure $SU(2)$ $\N=2$  super-Yang-Mills theory, as well as a generator associated to the non-Abelian locus in $\M_V$ that exchanges roughly $T_h$ with $S_h$  \cite{KlemmMayrK3,X8}.

From the heterotic viewpoint, the origin  of the $SU(2)$ gauge theory coupled to two adjoint hypermultiplets at $z_2=1$ is intrinsically non-perturbative and may be related to the existence of the NS5-brane. In fact, translated via S-duality into a type I picture \cite{het1}, NS5-branes  would be mapped into D5-branes that may play a role analogous to that of D1-branes already considered in Ref. \cite{Estes:2011iw}. There, D1-brane states winding internal 1-cycles were taken into account in the evaluation of the thermal free energy, whose effect was to stabilize internal moduli. Adding the contributions of D5-brane states winding internal 5-cycles in the evaluation of the free energy may lead to a stabilization of the type I dilaton.  Alternatively, the contributions of the solitonic D1-brane states were shown to be equivalently described in terms of E1-instantons wrapping the Euclidean time circle $S^1(R_0)$ and internal 1-cycles. Thus, it would be interesting to see if E5-instantons wrapping $S^1(R_0)$ and internal 5-cycles would contribute in such a way to generate a potential for the type I dilaton.

Finally, note that the depth of the minimum of the free energy density depends only on the number of classically massless states at this point. Therefore, the two minima in Eq. (\ref{inter}) are degenerate. Moreover, both are at finite distance in the compactified moduli space $\M_V$. Therefore, it may be interesting to find  instantonic transitions between them, and analyze resulting physical consequences.


\subsection{Discussion}
\label{discuss}

The qualitative behavior and stabilization issues of the example of compactification we have analyzed are shared by numerous models based on other families of threefolds, with small Hodge numbers $h_{11}$. For instance, cases where $h_{11}=2$ or 3, $N=2$ or 3 and $g=2,\dots, 15$ can be found in  Refs \cite{Katz,KlemmMyr}.

In fact, in any type II compactification on a CY space, we expect the vector multiplet moduli space $\M_V$ to be completely lifted once finite temperature is switched on, the latter point being certainly relevant to describe the cosmological evolution of our universe. Geometrically, this means that all \Ka moduli in type IIA and all complex structure moduli in type IIB have masses of order the temperature scale. From the IIA point of view (and similarly in the IIB mirror picture), the mechanism is based on the fact that all homology classes of 2-cycles can vanish and that D2-branes wrapped on their representatives should always give rise to non-perturbative BPS states that are massless (at zero temperature), when the cycles collapse.
In this work, this was analyzed in detail at conifold points, as well as at loci of $SU(N)$ enhanced gauge symmetries coupled to $g$ hypermultiplets in the adjoint representations. It would be interesting to extend our approach to other points where 2-cycles are vanishing, by identifying the geometrically engineered gauge theories  and the associated massless BPS states. For instance, one may analyze  the case of  non-Abelian gauge theories coupled to matter in the fundamental representations, which is considered in Ref. \cite{BKKM}.

Flat directions in the classical hypermultiplet moduli space $\M_H$ are also lifted. This is the case for the directions that realize branches of the above mentioned gauge theories. In other words, say in type IIA, the 3-cycles that can be resolved into 2-cycles when they collapse are expected to be associated to quaternionic directions in $\M_H$ lifted by the thermal effective potential. For instance, these scalars parameterize the Higgs branch arising at a conifold locus (when $R>S$), or the Coulomb or Higgs branches of the non-Abelian case we have analyzed. A question then arises: Can we stabilize this way all complex structure moduli in type IIA (\Ka moduli in type IIB)? To answer this question, let us consider a CY manifold $M$ admitting  2-cycles that cannot be deformed into 3-cycles. Such a case was understood physically in our study of the conifold locus in type IIA, when $R=S$ so that no Higgs branch exists. By mirror symmetry, there exist 3-cycles in the mirror CY $W$ that  cannot be resolved into 2-cycles. Utilizing $W$ to compactify the type IIA string, the eventual (gauge) theory realized geometrically in the vicinity of the vanishing locus in $\M_H$ of these 3-cycles is not know to us. Therefore, we are not able to identify possible massless states occurring at these points, which would induce a local minimum of the free energy density and a stabilization of the associated complex structure moduli of $W$. Clearly, it would be very interesting to clarify this issue.

The above discussion of $\M_H$ concerns the $h_{12}$ quaternionic directions associated to the complex structure of the internal space in type IIA. The remaining one, associated to the unique $(3,0)$-homology class, is parameterized by the scalars of the universal hypermultiplet, which contains the type II dilaton. Since the 3-cycles involved in the discussion of the tree level masses we considered can be resolved into 2-cycles, the universal hypermultiplet  was always ``spectator" and therefore unlifted by the  thermal effective potential. This fact is actually consistent with our restriction to the case of a dilaton field sitting in a weak coupling regime, throughout the  process of moduli stabilization. In fact, in string-frame, the one-loop correction to the vacuum energy is independent of the dilaton. In the Einstein frame, the vacuum energy acquires a dilaton dressing, which is however absorbed in the overall $T^4$ factor (see e.g. Eq. (\ref{PCL16})), where the temperature measured in this frame is defined as
\begin{align}
T={e^\phi \over 2\pi R_0}.
\end{align}
Therefore, it is only by taking into account higher loop corrections and non-perturbative effects that the thermal effective potential would source the type II dilaton, though at strong coupling. It may then be possible to study this regime in the dual heterotic picture, where the hypermultiplet moduli space $\M_H$ is exactly know, given the fact that $S_h$ sits in a vector multiplet. Working at heterotic weak coupling, the one-loop free energy evaluated on the heterotic side may stabilize the hypermultiplet dual to the universal one in type II.

In this paper, the attraction to a point $P_0$ in $\t\M_V\times \t\M_H$, where additional states become massless at zero temperature is shown provided the temperature is low enough compared to $M_{\rm min}$, the lower bound of the non-vanishing masses at $P_0$. Moreover, if at this point the $h_{11}$ homology classes vanish, it follows that $M_{\rm min}$ must be of order  $\O(e^\phi/\sqrt{\alpha'})$.  Therefore, the massive contributions $\O(e^{-M_{\rm min}/T})$ we neglected in the free energy (see e.g. Eq. (\ref{PCL16})) are exponentially suppressed, as soon as the universe exits the Hagedorn era and starts to cool.

At very early times, close to the Hagedorn temperature ($T\simeq e^\phi/\sqrt{\alpha'})$, the width of the potential $\F$ as a function of the moduli is very large. This follows from the fact that $e^{-M_s/T}$ (see Eq. (\ref{apG2})), where $M_s$ is a moduli-dependent mass that vanishes at $P_0$, is not narrow when $T$ is large.  Therefore, even if initially the moduli fields sit at a point $P$ very far from the local minimum at $P_0$, the well of the potential may contain both $P$ and $P_0$, so that the system is  dynamically attracted to a neighborhood of $P_0$, where the analysis of the present work starts to apply. Actually, the  well of the effective potential may overlap many CY moduli spaces, such as $M, M', M^\dprime, \dots$ connected by extremal transitions, so that the dynamical mechanism of moduli stabilization may favor CY compactifications with large Hodge numbers, for the local minima of $\F$ to be deep.


\section{Summary and perspectives}
\label{cl}
In this paper, we address the question of moduli stabilization in the context of type II superstring theory compactified on CY threefolds, once finite temperature is switched on. Even if the worldsheet conformal field theory is interacting, finite temperature can be implemented at the string level by a free orbifold action on the Euclidean time circle. This setup leads to no-scale models \cite{Noscale}, \ie classical theories where supersymmetry is spontaneously broken in flat Minkowski space. Therefore, flat directions of the classical potential exist, which can be organized as a product of  special \Ka and quaternionic manifolds, as follows from $\N=2$ supersymmetry.

The above moduli spaces admit particular loci, where the internal manifold develops singularities when 2-cycles or 3-cycles collapse, implying generically massive supermultiplets to become massless. For instance in type IIA, BPS D2-branes on vanishing 2-cycles lead to hypermultiplets charged under $U(1)$ factors at conifold loci \cite{Strominger:1995cz}, or $SU(N)$ enhanced gauge symmetries coupled to $g\ge 1$ hypermultiplets at some ``non-Abelian loci" \cite{Katz}. We show that at least in the weak string coupling regime, quantum/thermal effects stabilize the moduli at such particular points. The analysis is based on  the one-loop low energy effective action, {\em without integrating out} the above additional light states  in the sense of Wilsonian effective action,  in order to avoid any IR divergence. This perturbative analysis is justified by  the fact the gauge theories are non-asymptotically free. We first determine the classical part of the action, which is a supergravity theory, whose gauging induces a potential we use to determine the moduli-dependent classical masses of the extra light states. At one-loop and low enough temperature, the stringy Coleman-Weinberg effective potential depends on the classical masses and can be shown to admit local minima precisely where the light fields become classically massless.

The  scalars that are stabilized are those belonging to the vector multiplets and hypermultiplets involved in the gauge theories geometrically engineered in the vicinities of the loci, where the internal CY is singular. From the perspective of their stringy realizations, they can either be non-perturbative fields, or perturbative one, in which case they are identified with flat directions of the initial classical potential. Therefore, the mechanism stabilizes both \Ka and complex structure moduli. In fact, the points in moduli space that are favored are situated at the intersection of several loci, each of which being associated to singularities developed by the internal space. We have argued that in general, say in type IIA, the entire \Ka moduli space is expected to be lifted, as well as the complex structure moduli associated to 3-cycles which can be resolved into 2-cycles.

In this setup, the temperature $T$ is actually the no-scale modulus, also lifted by the thermal effective potential. However, instead of being stabilized (!) it acquires a run away behavior, which from a cosmological point of view arises when the flat universe expands. In other words, the model being non-supersymetric, time-translation is broken and the non-trivial one-loop contribution to the vacuum energy back-reacts on the classically static universe, which enters  in quasi-static
cosmological evolution. Homogenous and isotropic radiation dominated eras exist, characterized by static moduli sitting at their minima. They are particular solutions among more general ones, where the massive moduli oscillate with damping around their minima. However, their ``masses" happen to be proportional to the temperature, which is itself time-dependent and decreasing. As analyzed in detail in Ref. \cite{Estes:2011iw}, the energy density stored in their oscillations scales as $T^4$ rather than $T^3$, as is the case when scalars have constant masses. As a result, moduli never dominate over radiation and the cosmological moduli problem \cite{cosmomodprob} does not occur.

At one time or another, switching on finite temperature in a theoretical setup is certainly relevant to account for certain phases of the cosmological evolution of the universe. We stress that in the context of type II compactifications on CY threefolds, the effects described in this work should not be omitted. In the same process, they lead to stabilizations of moduli and determine the gauge group of the theory at low energy. In particular, the non-Abelian factors arise precisely at the points of enhanced gauge theory the moduli are attracted to.

However, more work is required to extend our results to compactifications on generalized CY spaces \cite{geneCY}, including fluxes, branes and orientifold projections, leading to $\N=1$ backgrounds at finite temperature. For $\N=1$ to remain broken when the temperature is low and recover an MSSM-like model, an additional source of  spontaneous breaking of supersymmetry should be implemented, whose origin may be attributed to the internal fluxes.  As explained in the introduction, it would be interesting to extend to this context the results of Refs \cite{CriticalCosmo,stabmod,dSi} derived in orbifold models. In these works, it is shown that  during an ``intermediate cosmological era", where the temperature evolves between the Hagedorn temperature and the electroweak scale, the time trajectories of the scale of spontaneous supersymmetry breaking $M(t)$ and temperature $T(t)$ are attracted to a particular solution, where they are proportional to the inverse scale factor, $M(t)\propto T(t)\propto 1/a(t)$. Therefore, as the universe expands and cools, the hierarchy $M\ll M_{Planck}$ is dynamically generated. In the process, the moduli acquire time dependent masses of order $M(t)^2/M_{Planck}$ or $M(t)$. At the end of the intermediate era, when the temperature reaches the electroweak scale, the radiative corrections of the soft supersymmetry breaking terms at low energy become large and are expected to induce the Higgs mechanisms. The latter is accompanied by the stabilization of the modulus $M(t)$ at some value of order the electroweak scale \cite{AlvarezGaume:1983gj,NoscaleTSR}.   It is at this epoch that the moduli masses should be definitively constant. After the above analysis worked out in details,  questions about dark matter may be addressed in this setup.

Finally, toroidal type II compactifications in presence of ``gravito-magnetic" fluxes lead to thermal models, free of Hagedorn-like divergences \cite{akpt}. The induced cosmological evolutions include bouncing \cite{dNonSingular} or emerging universes \cite{emer}, where no initial singularity is encountered, while remaining in a perturbative regime. Therefore, it would be interesting to see if gravito-magnetic fluxes can be implemented in (generalized) CY compactifications and possibly lead to a theoretical framework able to account for both very early and very late times cosmological eras.


\section*{Acknowledgments}

We are grateful to G. Bossard, E. Dudas, J. Estes, P. Fayet, A.-K. Kashani-Poor, A. Klemm, C. Kounnas, E. Palti and B. Pioline for fruitful discussions.
This work is partially supported by the EU contracts PITN GA-2009-237920, ERC-AG-226371 and IRSES-UNIFY, the French ANR 05-BLAN-NT09-573739 contract, the CEFIPRA/IFCPAR 4104-2 project, and PICS  contracts France/Cyprus, France/Greece and France/USA.


\section*{Appendix A:  't Hooft symbols}
\renewcommand{\theequation}{A.\arabic{equation}}
\renewcommand{\thesection}{A}
\setcounter{equation}{0}
\label{thooft}

We collect in this Appendix the definitions and  useful properties of the 't Hooft symbols. They are denoted ${\eta^{xu}}_v$ and ${\b \eta^{xu}}{}_v$ ($x=1,2,3$; $u,v=1,2,3,4$), are antisymmetric in $u$, $v$, and satisfy
\begin{align}
\label{thft}
{\eta^{xu}}_v={\b \eta^{xu}}{}_v={\epsilon^{xu}}_v\quad (u,v=1,2,3)\; ,\qquad {\eta^{xu}}_4=-\b \eta^{xu}{}_4=\delta^{xu},
\end{align}
where $\epsilon^{123}=1$. The indices $u$, $v$ are equally up or down, since they are raised or lowered by Kronecker symbols.
In matrix form, the 't Hooft symbols are written as
\begin{align}
\label{thft1}
    &\eta^1=\left(\!\! \begin{array}{cccc} 0& 0& 0& 1\\ 0& 0& 1& 0 \\0& -1& 0& 0\\-1& 0& 0& 0 \end{array} \!\!\right) ,\ \eta^2=\left(\!\! \begin{array}{cccc} 0& 0& -1& 0\\ 0& 0& 0& 1 \\1& 0& 0& 0\\ 0& -1& 0& 0 \end{array} \!\!\right),\ \eta^3=\left(\!\! \begin{array}{cccc} 0& 1& 0& 0\\ -1& 0& 0& 0 \\0& 0& 0& 1\\ 0& 0& -1& 0 \end{array} \!\!\right),\nonumber\\
    & \b\eta^1=\left(\!\! \begin{array}{cccc} 0& 0& 0& -1\\ 0& 0& 1& 0 \\0& -1& 0& 0\\1& 0& 0& 0 \end{array} \!\!\right)  ,\ \b\eta^2=\left(\!\! \begin{array}{cccc} 0& 0& -1& 0\\ 0& 0& 0& -1 \\1& 0& 0& 0\\ 0& 1& 0& 0 \end{array} \!\!\right),\ \b\eta^3=\left(\!\! \begin{array}{cccc} 0& 1& 0& 0\\ -1& 0& 0& 0 \\0& 0& 0& -1\\ 0& 0& 1& 0 \end{array} \!\!\right),
\end{align}
and fulfill the relations
\begin{align}
\label{thft2}
\eta^x\eta^y=-\delta^{xy}I_4-\epsilon^{xyz}\eta^z\; , \qquad\b\eta^x\b\eta^y=-\delta^{xy}I_4-\epsilon^{xyz}\b\eta^z,
\end{align}
which imply
\begin{align}
\label{thft3}
{\rm tr}\,\eta^x\eta^y={\rm tr}\, \b\eta^x\b\eta^y=-4\delta^{xy}.
\end{align}
Summing  over $x$, they give
\begin{align}
\label{thft4}
{\eta^{xt}}_u{\eta^{xv}}_w=\delta^{tv}\delta_{uw}-\delta^t_w\delta^v_u+{\epsilon^t}_u{{}^v}_w\; , \qquad {\b \eta^{xt}}{}_u{\b \eta^{xv}}{}_w=\delta^{tv}\delta_{uw}-\delta^t_w\delta^v_u-{\epsilon^t}_u{{}^v}_w\, ,
\end{align}
where $\epsilon^{1234}=1$.


\section*{Appendix B: Canonical basis in hypergeometry}
\renewcommand{\theequation}{B.\arabic{equation}}
\renewcommand{\thesection}{B}
\setcounter{equation}{0}
 \label{THEOREMS}

We first recall that the complex structures, metric and hyper-\Ka  forms defined on a vectorial space take simple forms, when they are written in a canonical base. We then apply these properties in the tangent plane at a given point $P_0$ of a quaternionic (or hyper-\K\"ahler) manifold. Finally, we find the canonical form of the Killing vectors associated to Abelian isometries that fix $P_0$.

\vskip .2cm
\noindent {\large \bf Theorem 1 :} Let $V$ be a $4n$-dimensional real vector space supplied with a triplet of complex structures $J^x$ ($x=1,2,3$), satisfying the quaternionic algebra
\begin{align}
\label{TRM0}
J^xJ^y=-\delta^{xy}\, I_{4n}+\epsilon^{xyz}J^z.
\end{align}
Let $V^*$ be the dual of $V$. Then, there always exists some basis $e_{\ca u}$ in $V$ and its dual $\theta^{\ca u}$ in $V^*$, where $\ca=1,\dots,n$ and $u=1,2,3,4$, such that the complex structures take the following form, in terms of 't Hooft symbols:
\begin{align}
\label{TRM1}
   J^x=-\delta^\A_\B\, \eta^{xu}{}_v\, e_{\ca u} \otimes \theta^{\cb v}.
\end{align}
Moreover, suppose $V$ is endowed with a metric $h$, which is Hermitian under the three complex structures $J^x$ ($x=1,2,3$),
\begin{align}
\label{hermi}
\forall v,w\in V: \quad h(J^xv,J^xw) \equiv h(v,w),
\end{align}
and define the triplet of hyper-\K\"ahler 2-forms $K^x$  by
\begin{align}
\label{KJ}
\forall v,w\in V:\quad K^x(v,w)\equiv h(J^xv,w).
\end{align}
Then, the basis $e_{\ca u}$ can always be chosen orthonormal,
\begin{align}
\label{hdiag1}
h=\delta_{\A\B}\, \delta_{uv}\, \theta^{\ca u}\otimes \theta^{\cb v},
\end{align}
and the hyper-\K\"ahler 2-forms take the canonical form
\begin{align}
K^x= \demi\, \delta_{\A\B}\, \eta^{x}_{uv}\, \theta^{\ca u}\wedge \theta^{\cb v}.
\end{align}


\noindent {\it \large Proof :}  Pick up any non-vanishing vector $e_{14}\in V$, and define $e_{1x}=-J^xe_{14}$. It is straightforward to consider an arbitrary linear combination of these four vectors to show that they are linearly independent. Denote $V_1=\text{Span}(e_{11},e_{12},e_{13},e_{14})$ and repeat the previous steps by taking a non-vanishing $e_{24}\in V\backslash V_1$. Apply $-J^x$ on it to define $e_{2x}$, and hence $V_2=\text{Span}(e_{21},e_{22},e_{23},e_{24})$. After repeating this procedure $n$ times, we obtain $V=V_1\oplus\dots \oplus V_n$. It is easy to check that in the basis $e_{\A u}$ ($\A=1,\dots,n$; $u=1,2,3,4$), the \Ka forms $J^x$ take the canonical form (\ref{TRM1}).

Next, defining $\theta^{\A u}$ to be the dual basis of $V^*$, we write the metric on $V$ as
\begin{align}
\label{TRM2}
h=h_{\ca u,\cb v}\, \theta^{\ca u}\otimes \theta^{\cb v},
\end{align}
and introduce the alternative notation $h_{uv}^{(\A\B)}=h_{\A u,\B v}$. Then, the Hermitian conditions (\ref{hermi}) lead to
\begin{align}
\label{TRM3}
[\eta^x,h^{(\A\B)}]=0.
\end{align}
This implies the real $4\times 4$ matrix $h^{(\A\B)}$ can be written as
\begin{align}
\label{TRM8}
    h^{(\A\B)}=h^{(\B\A)T}=\left(\begin{array}{rrrr} a^{(\ca\cb)} & b^{(\ca\cb)} & c^{(\ca\cb)} & d^{(\ca\cb)} \\ -b^{(\ca\cb)} & a^{(\ca\cb)} & -d^{(\ca\cb)} & c^{(\ca\cb)} \\ -c^{(\ca\cb)} & d^{(\ca\cb)} & a^{(\ca\cb)} & -b^{(\ca\cb)} \\ -d^{(\ca\cb)} & -c^{(\ca\cb)} & b^{(\ca\cb)} & a^{(\ca\cb)} \end{array}\right)
\end{align}
and satisfies
\begin{align}
\label{TRM5}
 h^{(\ca\cb)}h^{(\cb\ca)}=\lambda^{(\ca\cb)}\, I_4 \quad \where \quad \lambda^{(\ca\cb)}=(a^{(\ca\cb)})^2+(b^{(\ca\cb)})^2+(c^{(\ca\cb)})^2+(d^{(\ca\cb)})^2.
\end{align}
In particular, $h^{(\A\A)}$ is diagonal with $a^{(\A\A)}>0$, for the metric $h$ to be definite positive. Thus, we can always rescale $e_{\ca u}$ to effectively set $h^{(\A\A)}=I_4$. Clearly, such a rescaling does not spoil Eq. (\ref{TRM1}). Now, we introduce a scheme that removes the off-diagonal blocks of the metric, $h^{(\A\B)}$ for $\A\neq \B$, while keeping the standard form of the complex structures. We work this out block by block.

Take $(\ca,\cb)=(1,2)$, and exhibit the relevant part of the metric as
\begin{align}
h=(\theta^{1T},\theta^{2T})\left(\!\!\begin{array}{cc}I_4 & h^{(12)}\\ h^{(21)} &I_4\end{array}\!\!\right)\left(\!\!\begin{array}{c}\theta^1\\ \theta^2\end{array}\!\!\right) +\cdots.
\end{align}
It is straightforward to see that under the  change of dual basis of $V^*$ and $V$,
\begin{align}
\left(\!\!\begin{array}{c}\theta^1\\ \theta^2\end{array}\!\!\right)=\left(\!\!\begin{array}{cc}I_4 & -h^{(12)}\\ 0_4 &I_4\end{array}\!\!\right)\left(\!\!\begin{array}{c}\theta^{\prime1}\\ \theta^{\prime2}\end{array}\!\!\right), \quad (e_1,e_2)=(e'_1,e'_2)\left(\!\!\begin{array}{cc}I_4 & h^{(12)}\\ 0_4 &I_4\end{array}\!\!\right),
\end{align}
the metric becomes
\begin{align}
\label{hdiag}
h=(\theta^{\prime1T},\theta^{\prime2T})\left(\!\!\begin{array}{cc}I_4 & 0_4\\ 0_4 &(1-\lambda^{(12)})I_4\end{array}\!\!\right)\left(\!\!\begin{array}{c}\theta^{\prime1}\\ \theta^{\prime2}\end{array}\!\!\right) +\cdots,
\end{align}
while the canonical form of the complex structures (\ref{TRM1}) is conserved, as follows from Eq. (\ref{TRM3}). Since Eq. (\ref{hdiag}) and the positive-definiteness of $h$ imply $\lambda^{(12)}<1$, it is possible to rescale the $\theta^{\prime2u}$'s to absorb the factor $(1-\lambda^{(12)})$ in this equation. Again, this operation conserves the form of the complex structures.

Then, we apply the same procedure to eliminate the block $(\A,\B)=(1,3)$. It is easily seen that this procedure does not reintroduce a non-trivial bloc $(1,2)$. In general, one can show by double recursion that for $\A=1,\dots,n-1$ and $\B=\A+1,\dots,n$, one can get rid off the blocks $h^{(\A\B)}$. At the end of this process, the metric is diagonalized, $h=\theta^{\ca u} \theta^{\ca u}$, and Eq. (\ref{TRM1}) is valid.
In this basis, the components of the hyper-\K\"ahler forms are
\begin{align}
K^x_{\A u,\B v}=h_{\B v,\C w}J^{x\C w}{}_{\A u}=\delta_{\A\B}\, \eta^x_{uv}.
\end{align}


\vskip .2cm
\noindent {\large \bf Theorem 2 :} Let $\cm$ be a quaternionic (or hyper\Ka) manifold of dimension $4n$, and any given point $P_0\in\M$. Then, there exists some local coordinates $q^{\ca u}$ ($\ca=1,\dots,n; u=1,2,3,4$) such that  $\left. q^{\ca u}\right\abs_{P_0}=0$ and the complex structures $J^x$, the metric $h$ and the hyper-\K\"ahler forms $K^x$ at $P_0$ are:
\begin{align}
\label{cond1}&\left.J^x\right|_{P_0}=-\eta^{xu}{}_v\left.\Big({\partial \over \partial q^{\ca u}}\otimes dq^{\ca v} \Big)\right\abs_{P_0}, \\
\label{cond2}&\left.h\right|_{P_0}= \left.\big(dq^{\ca u}dq^{\ca u} \big)\right\abs_{P_0}, \\
\label{cond3}&\left.K^x\right|_{P_0}=\demi \eta^x_{uv}\left.\big(dq^{\ca u} \wedge dq^{\ca v} \big)\right\abs_{P_0}.
\end{align}


\noindent {\it \large Proof :}  Consider a chart $\{\U,q^{\Lambda}\}$, where $\U$ is an open neighborhood of $P_0$ and $q^\Lambda$ some coordinate system in $\U$. Let $q_0^\Lambda$ be the coordinates of $P_0$. Applying Theorem 1 to the tangent plane at any point $P\in \U$, there exists a vielbein $\theta^{\ca u}$ and its dual $e_{\ca u}$ in $\U$, such that
\begin{align}
J^x=-\eta^{xu}{}_v\, e_{\A u}\otimes\theta^{\A v}, \quad h=\theta^{\A u}\otimes\theta^{\A u}, \quad K^x={1\over 2}\,\eta^x_{uv}\, \theta^{\A u}\wedge\theta^{\A v}.
\end{align}
We can write $\theta^{\ca u}= U^{\ca u}{}_{\Lambda}\, dq^{\Lambda}$ and $e_{\ca u}= U^{-1\Lambda}{}_{\ca u}\, {\dis \partial\over \dis \partial q^{\Lambda}}$, where the matrix $\big( U^{\ca u}{}_\Lambda \big)$ is invertible and depends smoothly on $P\in \U$. The new coordinates in $\U$
\begin{align}
q^{\A u}:= \left.U^{\ca u}{}_\Lambda\right\abs_{P_0}\, (q^\Lambda-q^\Lambda_0)
\end{align}
satisfy Eqs (\ref{cond1})--(\ref{cond3}) and vanish at $P_0$.


\vskip .2cm
\noindent  {\large \bf Abelian isometries :} In order to describe the charged  hypermultiplets sector of an Abelian gauge theory, we suppose from now on the manifold $\M$ in Theorem 2 admits $U(1)^S$ isometries with fixed point $P_0$. Our aim is to find a canonical form for the Killing vectors at $P_0$.

We know from the first part of Theorem 1 (see Eqs (\ref{TRM0}) and (\ref{TRM1})) applied to the tangent plane at $P_0$ that there is a system of coordinates $q^{\A u}$ in the neighborhood $\U$ of $P_0$ such that (\ref{cond1}) is satisfied an $\left.q^{\A u}\right\abs_{P_0}=0$. We are interested in metrics on $\M$ admitting $U(1)^S$ isometries, whose Killing vectors have components $k_i^{\A u}$ ($i=1,\dots, S$) admitting Taylor expansions of the form
\begin{align}
\label{kil}
k^{\A u}_i=Q_i^{\A u}\, t^u{}_v\, q^{\A u}+\O(q^2).
\end{align}
By construction, the point $P_0$ is fixed under the action of $U(1)^S$. Moreover, the isometries do not mix the components of different hypermultiplets. In other words, the quadruplet $(q^{\ca 1}, q^{\ca 2}, q^{\ca 3}, q^{\ca 4})$ has a well defined charge $Q^\A_i$ under the $i^{\rm th}$ $U(1)$. The $U(1)$ generator $ t^u{}_v$ in Eq. (\ref{kil}) is determined by the convention  to define complex numbers in the affine plane at $P_0$. For instance, the multiplication by the imaginary number $i$ is represented by  $J^3$, when we combine the $q^{\A u}$'s into complex numbers $q^{\ca 1}+iq^{\ca 2}$ and $q^{\ca 3}+iq^{\ca 4}$. In this case, the infinitesimal $U(1)^S$ transformations
\begin{align}
e^{i\epsilon^iQ_i^\A}\, (q^{\ca 1}+i q^{\ca 2})=(q^{\ca 1}+i q^{\ca 2})+\epsilon^iQ_i^\A
(-q^{\ca 2}+i q^{\ca 1})+\cdots,\no\\
e^{-i\epsilon^iQ_i^\A}\, (q^{\ca 3}+i q^{\ca 4})=(q^{\ca 3}+i q^{\ca 4})+\epsilon^iQ_i^\A
(q^{\ca 4}-i q^{\ca 3})+\cdots,
\end{align}
we want to represent with $\delta q^{\A u}=\epsilon^ik_i^{\A u}$ imply $t=-\bar\eta^3$.

A question then arises. Is the first order form of the Killing vector (\ref{kil}) conserved, when we diagonalize $h$, while keeping the canonical form of the complex structures $J^x$ at $P_0$? The answer to this question is yes, due to the fact that the metric $h$ must satisfy Killing equation
\begin{align}
\label{QTG5}
h_{\A u,\C w}\, D_{\B v} k_i^{\C w}+h_{\B v,\C w}\, D_{\A u}k_i^{\C w}=0,
\end{align}
where $D$ denotes the covariant derivative on $\M$. At order zero in $q^{\A u}$, this relation becomes 
\begin{align}
\label{QTG5'}
\left.h_{\A u,\B w}\right\abs_{P_0} Q^{\B}_i\, t^u{}_v+\left.h_{\B v,\A w}\right\abs_{P_0}\, Q^{\A}_i\, t^w{}_u=0.
\end{align}
To diagonalize the hermitian metric $h$ without spoiling the form of Eq. (\ref{cond1}), we saw in the proof of Theorem 1 that we can set the blocks $h^{(\A\A)}=I_4$ and eliminate successively all non-diagonal $4\times 4$ blocks $h^{(\A \B)}$ ($\A\neq \B$) of the metric by performing a sequence of changes of basis. These changes of bases for blocks $h^{(\A \A)}$ are only rescalings of the $q^{\A u}$'s which certainly conserve the form of the first order expansion of the Killing vectors $k_i^{\C w}\partial_{\C w}$. For the blocks $h^{(\A \B)}$ with $\A < \B$, they are of the form
\begin{align}
\left(\!\!\begin{array}{c}q^\A\\ q^\B\end{array}\!\!\right)=\left(\!\!\begin{array}{cc}I_4 & \left.-h^{(\A\B)}\right\abs_{P_0}\\ 0_4 &I_4\end{array}\!\!\right)\left(\!\!\begin{array}{c}q^{\prime\A}\\ q^{\prime\B}\end{array}\!\!\right), \quad \Big({\partial\over \partial q^\A},{\partial\over \partial q^\B}\Big)=\Big({\partial\over \partial q^{\prime\A}},{\partial\over \partial q^{\prime\A}}\Big)\left(\!\!\begin{array}{cc}I_4 & \left.h^{(\A\B)}\right\abs_{P_0}\\ 0_4 &I_4\end{array}\!\!\right),
\end{align}
which conserve the first order form of $k_i^{\C w}\partial_{\C w}$ as well, as can be checked using Eq. (\ref{QTG5'}).
To summarize, we have shown that there exists a system of coordinates $q^{\A u}$ on $\M$ such that $\left.q^{\A u}\right\abs_{P_0}=0$ and Eqs (\ref{cond1})--(\ref{cond3}) and (\ref{kil}) are statisfied.


\section*{Appendix C}
\renewcommand{\theequation}{C.\arabic{equation}}
\renewcommand{\thesection}{C}
\setcounter{equation}{0}

\noindent {\large \bf Theorem 3 :}
Let $\M$ be a quaternionic manifold of dimension $4n$ and $\omega^x$ the connection of the associated $SU(2)$ principal bundle.  The fiber bundle over $\M$, whose fibers are triplets of $SU(2)$, does not admit non-trivial local parallel sections. In other words, the equation
\begin{align}
 \nabla^{SU(2)}L^x\equiv dL^x+\epsilon^{xyz}\omega^yL^z=0
\label{TRM12}
\end{align}
in an open subset $\U$ of $\M$ has only the solution $L^x=0$.


\noindent {\it \large Proof :}
We carry out a point-wise proof. For any given point $P_0\in \U$, we consider the coordinate system $q^{\A u}$ of Theorem 2 and write Eq. (\ref{TRM12}) as
\begin{align}
\forall P\in\U:\qquad {\partial L^x\over \partial q^{\A u}}+\epsilon^{xyz}\omega^y_{\ca u} L^z=0.
\end{align}
Taking the partial derivative $\partial/\partial q^{\B v}$ and antisymmetrizing in $(\ca u, \cb v)$ yields
\begin{align}
\epsilon^{xyz}\Omega^y_{\ca u,\cb v} L^z=0,
\end{align}
where $\Omega^x=d\omega^x+\demi \epsilon^{xyz} \omega^y\wedge \omega^z$ is the curvature 2-form of the $SU(2)$-bundle.
Since $\M$ is quaternionic, we have $\Omega^x=\lambda K^x$,  where $\lambda$ is a non-vanishing constant. Given the fact that $\left.K^x\right|_{P_0}$ satisfies Eq. (\ref{cond3}), we obtain $\epsilon^{xyz}\eta^y_{uv} \left.L^z\right|_{P_0}=0$. Multiplying with $\eta^{xw}{}_u$ and summing over $x$ and $u$, Eq. (\ref{thft2}) leads to  $\left.\eta^{zw}{}_vL^z\right|_{P_0}=0$.
Multiplying with $\eta^{xv}{}_w$ and summing over $w$ and $v$, our desired result $\left.L^x\right|_{P_0}=0$ is obtained using Eq. (\ref{thft3}).



\end{document}